\documentclass[10pt, reqno]{amsart}

\usepackage[utf8]{inputenc}
\usepackage{fontenc}
\usepackage{graphicx}
\usepackage{bbm}
\usepackage [centertags] {amsmath}
\usepackage {amsfonts}
\usepackage{amsmath}
\usepackage{amssymb}
\usepackage {amsthm}
\usepackage {amscd}
\usepackage{enumerate}
\usepackage{cite}

\newtheorem{theorem}{Theorem}[section]
\newtheorem{lemma}[theorem]{Lemma}
\newtheorem{corollary}[theorem]{Corollary}

\newtheorem{proposition}[theorem]{Proposition}

\newcommand{\HilbertSp}{\mathcal{H}}

\newcommand{\transpos}{t}
\newcommand{\coset}[1]{#1^{\tau}}
\newcommand{\cosetb}[1]{\coset{\left(#1\right)}}

\newcommand{\kpositivitymaps}[1]{\mathcal{P}_{#1}}

\newcommand{\kspositivitymaps}[1]{\mathcal{SP}_{#1}}
\newcommand{\Kapositivemaps}[1]{\mathcal{P}_{#1}}
\newcommand{\Kapositivemapsb}[2]{\Kapositivemaps{#1}\left(#2\right)}
\newcommand{\PAK}[2]{P\left(#1,#2\right)}

\newcommand{\kmdecmaps}[2]{\mathcal{D}_{#1,#2}}
\newcommand{\kmpositivitymaps}[2]{\mathcal{P}_{#1,#2}}
\newcommand{\kmspositivitymaps}[2]{\mathcal{S}_{#1,#2}}
\newcommand{\kmdecmapsb}[3]{\kmdecmaps{#1}{#2}\left(#3\right)}
\newcommand{\kmpositivitymapsb}[3]{\kmpositivitymaps{#1}{#2}\left(#3\right)}
\newcommand{\kmspositivitymapsb}[3]{\kmspositivitymaps{#1}{#2}\left(#3\right)}

\newcommand{\posops}{B^+}
\newcommand{\bposops}{BP}
\newcommand{\kbposops}[1]{#1\textrm{-}BP}
\newcommand{\sepops}{Sep}
\newcommand{\ksepops}[1]{#1\textrm{-}Sep}
\newcommand{\linearmapsb}[1]{\mathcal{L}\left(#1\right)}
\newcommand{\hermicitymapsb}[1]{\mathcal{E}\left(#1\right)}
\newcommand{\positivitymapsb}[1]{\mathcal{P}\left(#1\right)}
\newcommand{\spositivitymapsb}[1]{\mathcal{SP}\left(#1\right)}
\newcommand{\kpositivitymapsb}[2]{\kpositivitymaps{#1}\left(#2\right)}
\newcommand{\cpositivitymapsb}[1]{\mathcal{CP}\left(#1\right)}
\newcommand{\kspositivitymapsb}[2]{\kspositivitymaps{#1}\left(#2\right)}
\newcommand{\linearopsb}[1]{B\left(#1\right)}
\newcommand{\hermitianopsb}[1]{E\left(#1\right)}
\newcommand{\posopsb}[1]{B^+\left(#1\right)}
\newcommand{\bposopsb}[1]{BP\left(#1\right)}
\newcommand{\kbposopsb}[2]{#1\textrm{-}BP\left(#2\right)}
\newcommand{\sepopsb}[1]{Sep\left(#1\right)}
\newcommand{\ksepopsb}[2]{#1\textrm{-}Sep\left(#2\right)}

\newcommand{\Choimatr}[1]{C_{#1}}
\newcommand{\Jamiosymb}{J}
\newcommand{\Jamiso}[1]{\Jamiosymb\left(#1\right)}
\newcommand{\dualitymaps}[1]{#1^{\circ}}
\newcommand{\dualityops}[1]{#1^{\circ}}
\newcommand{\dualitymapsb}[1]{\left(#1\right)^{\circ}}
\newcommand{\dualityopsb}[1]{\left(#1\right)^{\circ}}
\newcommand{\hconj}[1]{#1^{\ast}}
\newcommand{\hconjb}[1]{\left(#1\right)^{\ast}}

\newcommand{\cconjb}[1]{\left(#1\right)^{\ast}}
\newcommand{\innerpr}[2]{\left<#1\right|\left.#2\right>}
\newcommand{\HSprod}[2]{#1\cdot #2}
\newcommand{\diad}[2]{\left|#1\right>\left<#2\right|}
\newcommand{\proj}[1]{\diad{#1}{#1}}
\newcommand{\identitymap}{\mathbbm{1}}
\newcommand{\identitymapn}[1]{\identitymap_{#1}}
\newcommand{\setC}{\mathbbm{C}}
\newcommand{\setR}{\mathbbm{R}}
\newcommand{\setN}{\mathbbm{N}}
\def\Tr{\mathop{\textnormal{Tr}}}

\def\rk{\mathop{\textnormal{rk}}}
\def\Ad{\textnormal{Ad}}
\newcommand{\Tra}[1]{\Tr{#1}}
\newcommand{\Trb}[1]{\Tr{\left(#1\right)}}
\newcommand{\seq}[3]{\left\{#1\right\}_{#2}^{#3}}
\newcommand{\SEQ}[3]{$\left\{#1\right\}_{#2}^{#3} $}
\newcommand{\convhull}[1]{\mathop{\textnormal{convhull}}#1}
\newcommand{\Hspdimension}{d}
\newcommand{\semiposdef}{positive}
\newcommand{\adjacencymap}{similarity map}
\newcommand{\parttr}{\identitymap\otimes\transpos}
\newcommand{\parttrb}[1]{\left(\identitymap\otimes\transpos\right)#1}
\newcommand{\cstar}{C^{\ast}}
\newcommand{\cstardollar}{$\cstar$}

\begin{document}
\title{Cones of positive maps and their duality relations}
\author{{\L}ukasz Skowronek, Erling St{\o}rmer, Karol {\.Z}yczkowski}
\address{Institute of Physics\\
Jagiellonian University\\
30-059 Krakow, Poland}
\email{lukasz.skowronek@uj.edu.pl}
\address{ Department of Mathematics\\ 
	University of Oslo\\
	0316 Oslo, Norway}
\email{erlings@math.uio.no}
\address{ Institute of Physics\\ 
	Jagiellonian University\\
	30-059 Krakow, Poland
	\vskip 0.1 mm
	Center for Theoretical Physics\\
	PAN\\
	02-668 Warszawa, Poland
}
\email{karol@cft.edu.pl}

%

\begin{abstract}
The structure of cones of positive and  $k$-positive maps 
acting on a finite-dimensional Hilbert space is investigated.
Special emphasis is given to their duality relations to the sets of superpositive and $k$-superpositive maps.
We characterize $k$-positive and $k$-superpositive maps with regard to their properties under taking compositions.
A number of results obtained for maps are also rephrased for the corresponding cones of 
block positive, $k$-block positive,
separable and $k$-separable operators, due to the Jamio{\l}kowski-Choi isomorphism.
Generalizations to a situation where no such simple isomorphism is available are also made, employing the idea of mapping cones.
As a side result to our discussion, we show that extreme entanglement witnesses,
which are optimal, should be of special interest in entanglement studies.
\end{abstract}
\maketitle

\section{Introduction}

Positive linear maps of \cstardollar-algebras has been a subject of the mathematical literature for 
several years. In short, such a map sends the cone of positive operators acting on 
a given Hilbert space into itself. A map  $\Phi$ is called \textit{completely positive} (CP),
if the tensor product $\Phi \otimes\identitymapn{k}$ is positive for any 
dimension $k$ of an auxiliary Hilbert space.

On the one hand, the structure of the set of completely positive maps,
which forms a proper subset of the set of positive maps, is already well understood.
Completely positive maps find direct application in quantum theory as they correspond 
to \textit{quantum operations}, which can be realized in a physical experiment. 
On the other hand, in spite of a considerable effort several years ago 
\cite{St63,dePillis,Av69,Ja72,Ch72,Cho75,Wo76,Wo76b,St82,TT83,Ro85, St86,Os91,CKL92}
and more recently \cite{Ha98,EK00,MM01,Ko03,Ha03,Ky03,BFP04,CK07,St08,St09}
the structure of the set of positive maps acting on operators defined on a 
$d$ dimensional Hilbert space ${\mathcal H}_d$ is well understood only for $d=2$.
In this case every positive map is decomposable, as it can be represented
as a sum of a completely positive map and a completely co-positive map.

This mathematical fact, following from the results of St{\o}rmer \cite{St63} 
and Woronowicz \cite{Wo76},
has profound 
consequences for the entire theory of quantum entanglement.
It implies that the commonly used PPT criterion for quantum separability \cite{Pe96}
works in both directions only for $2 \times 2$ quantum systems \cite{HHH96a}. 
In other words, any state of a two qubit system is separable if and only if
it has the property of positive partial transpose (PPT).
Hence in this simplest case the sets of separable states and PPT states coincide,
and any state characterized by a negative partial transpose is entangled.

This is not the case for higher dimensions. For instance, 
the existence of non-decomposable positive maps shown for $d=3$ 
 by Choi \cite{Cho75}, implies that for a $3 \times 3$ quantum system
there exist PPT entangled states. Such quantum states are called
\textit{bound entangled} \cite{HHH98}, 
as they cannot be distilled into maximally entangled states,
and their subtle properties became recently a subject of 
a vivid scientific interest \cite{HHH99,HHHH08}.
In general, the question of characterizing the set of entangled
states for an arbitrary quantum system composed of two subsystems of size 
$d$, remains as one of the key  unsolved problems in the theory 
of quantum information. However, from a mathematical perspective
this problem is related to characterization of the set of all
positive maps in $d$ dimensions, which is known to be difficult.

It is convenient to define a subclass of positive maps, called \textit{$k$-positive}, such
that  $\Phi \otimes {\mathbbm 1}_k$ is positive.\footnote{Let us emphasize here the difference between 
$k$-positive maps defined for an integer $k$ 
and $\mathcal K$-positive maps \cite{St86,St09}, in which  $\mathcal K$ denotes a certain cone of operators.}
It is well known that $d$-positive maps are completely positive \cite{Cho75a}. 
Due to the theorem of Stinespring \cite{St55} any CP map can be represented as a sum
of \textit{\adjacencymap s}: $x \mapsto x_i:= a_i^* x  a_i$, 
where $^*$ denotes the Hermitian conjugation, and the operators $a_i$ are arbitrary.
In physics literature the operators $a_i$ are called \textit{Kraus operators}, \cite{Kr71} 
and it is possible to find such representation
for which the number of them does not exceed $d^2$.

In general the operators $a_i$ are of rank $d$, but it is useful to distinguish 
the class of linear maps for which there exists a representation into 
Kraus operators of rank not greater than $k$, where $k=1,\dots,d-1$.
These maps will be called \textit{$k$-superpositive}, 
since in the case $k=1$, the set of maps (denoted by $S\left(H\right)$ in \cite{St86})
for which all Kraus operators can be chosen to be of rank $1$, 
coincides with the set of \textit{superpositive maps}, introduced by Ando \cite{An04} (see also \cite{HSW05}).

Any linear map acting on a set of positive operators on 
${\mathcal H}_d$ represents an operator acting on the
composed Hilbert space ${\mathcal H}_d \otimes {\mathcal H}_d$.
This fact, known as \textit{Jamio{\l}kowski}
isomorphism due to his early contribution \cite{Ja72},
implies an intrinsic relation between the sets
of quantum maps and quantum states \cite{ZB04,AKMS06}.
In particular, positive maps correspond to \textit{block positive} operators \cite{Ja72},
while completely positive maps are represented by positive operators \cite{Cho75a}.
Thus a positive matrix representing  a completely positive map in this isomorphism
is called a \textit{Choi matrix} or  \textit{dynamical matrix} \cite{SMR61}.

Making use of the standard Hilbert-Schmidt scalar product of two operators
one can introduce a duality relation between sets of operators.
The set of positive operators $\posops$ is selfdual. 
The sets of block positive operators is known to be dual to the set
of  separable operators.
Therefore we cannot resists a temptation to call elements of the 
set dual to the set of \textit{$k$-block positive} operators
as \textit{$k$-separable}, although the same set appears in the literature
\cite{TH00,SBL01,HBLS04} and is characterized by the maximal 
Schmidt number of its element.

Note that the sets of operators which are a) block positive,
b) $2$-block positive, c) positive, d) $2$-separable and e) separable,
form a nested chain of proper subsets, see Fig. \ref{figuredualities0} and Table \ref{table1cones}
The same inclusion relations holds for the corresponding sets
of maps. As the elements of the cone dual to the cone of
positive maps are called  superpositive maps \cite{An04} 
(or \textit{entanglement breaking channels} \cite{HSR03,HSW05}),
the dual to the set of $k$-positive maps
consists of $k$-superpositive maps.

Since the set of block positive operators and separable operators
are dual, any positive map (which is not completely positive)
can be used to detect quantum entanglement.
In particular, the Choi matrix representing such a map
is given by a block positive operator
and it may play the role of an \textit{entanglement witness} \cite{HHH96a,Te00}.

\begin{table}[htp]
\begin{tabular}
[c]{|c|c|c|c|c|}\hline
& \multicolumn{2}{|c|}{Linear maps} & \multicolumn{2}{|c|}{Operators acting on ${\mathcal {H}}_d \otimes {\mathcal {H}}_d$}
\\\cline{2-5}\cline{2-5}%
$k$ & a) cone & b) dual cone & a') cone  & b') dual
cone\\\hline \hline
$1$ & positive  & superpositive & block positive  & separable\\\hline
$2$ & $2$-positive & $2$-superpositive & $2$-block positive & $2$-separable
\\\hline
$...$ & $...$ & $...$ & $...$ & $...$\\\hline
\!$d\!-\!1$\! & $(d\!-\!1)$-positive & \!$(d\!-\!1)$-superpositive \!& \!$(d\!-\!1)$-block positive \! &
$\!(d\!-\!1)$-separable\! \\\hline
$d$ & \multicolumn{2}{|c}{completely positive} &
\multicolumn{2}{|c|}{positive}
\\\hline
\end{tabular}\vskip 3 mm
\caption{\label{table1cones} The cones of linear maps acting on the set of 
operators on ${\mathcal {H}}_d$  and the isomorphic cones of operators.
Strict inclusion relations 
 hold upwards ($\cup$) for the cones in columns a), a') and
downwards ($\cap$) for the corresponding dual cones in columns b) and b'). In the case $k=\Hspdimension$ the cone of
completely positive maps is selfdual and so is the 
corresponding cone of positive operators.}
\end{table}

Such a hermitian operator $W$ is characterized by the property 
that Tr$W\sigma\geqslant 0$ for any separable state $\sigma$,
while negativity of Tr$W\rho $
confirms that the analyzed state $\rho$ is entangled.
The key advantage of this notion is due to the fact
that the Hermitian operator  $W$ can be considered as an
observable, and the expectation value
Tr$W\rho $ can be decomposed into a sum of quantities, 
which may be directly \textit{measured} in a laboratory.
In such a way one may experimentally
confirm that an analyzed quantum state $\rho$ is 
indeed entangled \cite{GHB+03,MEK+04}.

The set of entanglement witnesses corresponds thus to the
set of block positive operators, the structure of which
for $d \geqslant 3$ is still being investigated \cite{LKCH00,CK07,Sa08}. 
It is worth to emphasize that there is no universal witness,
which could detect entanglement of any state,
but for any entangled state a suitable witness can be found.
The most valuable are extreme entanglement witnesses, which 
form extreme points of the set of block positive operators,
since they can also detect entanglement of some weakly entangled states.
In this way the theory of quantum information provides a direct
motivation to study the structure of the set of block positive 
operators (i.e. the set of entanglement witnesses) and its
various subsets.

The aim of this work is to contribute to
understanding of the non trivial structure of the set
of positive maps and the corresponding set of 
block positive operators.
We provide a constructive characterization of
various subsets of the set of positive maps.
In particular we study relations based on duality
between convex cones. Another class of results
concerns composition of quantum maps.

This paper is organized as follows.
In section 2 we review necessary definitions 
of $k$-positive and $k$-superpositive maps
and formulate a kind of generalized 
Jamio{\l}kowski-Choi theorem, which
relates them to $k$-block positive
and $k$-separable operators.
Several other characterizations
of these sets are proved.
In section 3 we discuss the duality 
between the cones of $k$ positive and
$k$-superpositive maps and analyze its
consequences. 

In section \ref{secmappingcones} we study the relations of the results
obtained in the previous sections to $\mathcal K$-positive maps, where $\mathcal K$ is a so-called \textit{mapping cone}, introduced in \cite{St86}.


\section{Cones of positive maps and the corresponding sets of operators}

In this section we give the definitions 
to which we refer in later parts of the paper
and provide some concrete examples of objects that match these definitions.
We review certain results already known in the literature
and for convenience of the reader we prove some of them.

In the entire paper, we shall consider only finite dimensional linear spaces. 
Let $\HilbertSp={\mathcal H}_d$ be a Hilbert space of finite dimension $\Hspdimension$.
We denote by $\linearopsb{\HilbertSp}$ ($\hermitianopsb{\HilbertSp}$, 
$\posopsb{\HilbertSp}$) the set of linear (resp. hermitian, \semiposdef) 
operators on $\HilbertSp$. We choose an orthonormal basis 
\SEQ{e_i}{i=1}{\Hspdimension} of $\HilbertSp$ and the corresponding complete set of matrix units
\SEQ{e_{ij}}{i,j=1}{\Hspdimension} in $\linearopsb{\HilbertSp}$.

Let us consider the set $\linearmapsb{\HilbertSp}$ of linear 
maps sending $\linearopsb{\HilbertSp}$ into itself. 
An element $\Phi$ of $\linearmapsb{\HilbertSp}$ is called 
\textit{Hermiticity preserving} iff 
$\Phi\left(\hermitianopsb{\HilbertSp}\right)\subset\hermitianopsb{\HilbertSp}$. 
\textit{Positive maps} are the elements $\Phi$ which fulfill 
$\Phi\left(\posopsb{\HilbertSp}\right)\subset\posopsb{\HilbertSp}$. 
The set of Hermiticity preserving maps will be denoted by 
$\hermicitymapsb{\HilbertSp}$ and the set of positive maps by 
$\positivitymapsb{\HilbertSp}$. It is easy to show (cf. \cite{MastersThesis}) that 
positivity of $\Phi\in\linearmapsb{\HilbertSp}$ implies the Hermiticity preserving 
property, so we have the inclusion 
$\positivitymapsb{\HilbertSp}\subset\hermicitymapsb{\HilbertSp}$. Let $k$ be a 
positive integer. The family of \textit{$k$-positive maps}, 
$\kpositivitymapsb{k}{\HilbertSp}$, is defined by the condition 
$\identitymapn{k}\otimes\Phi\in\positivitymapsb{\setC^k\otimes\HilbertSp}$. 
That is, $\Phi\in\linearmapsb{\HilbertSp}$ is $k$-positive iff the tensor product of 
$\Phi$ by the $k$-dimensional identity map $\identitymapn{k}$ remains positive. 
A different characterization of $k$-positivity is given in the following lemma,
\begin{lemma}\label{lemmakposchar}
Let $\Phi$ be an element of $\linearmapsb{\HilbertSp}$. The map $\Phi$ is 
$k$-positive iff the map
\begin{equation}
\label{kposAd}
\linearopsb{\HilbertSp\otimes\HilbertSp}\ni x\longmapsto\left(\identitymapn{\Hspdimension}
\otimes\Phi\right)\left(q\otimes\identitymapn{d}\right)x\left(q\otimes\identitymapn{d}\right)\in\linearopsb{\HilbertSp\otimes\HilbertSp}
\end{equation}
is positive for an arbitrary $k$-dimensional orthogonal projection $q$ in $\HilbertSp$.
\begin{proof}
 Let $q=\sum_{i=1}^k\proj{f_i}$, where $\seq{f_i}{i=1}{\Hspdimension}$ is an
 orthonormal basis of $\HilbertSp$. We choose 
$\seq{f_i\otimes e_j}{i,j=1}{\Hspdimension}$ as the orthonormal basis of 
$\HilbertSp\otimes\HilbertSp$. The map \eqref{kposAd} is positive iff it is positive
 on all one dimensional projections on $\HilbertSp\otimes\HilbertSp$,
\begin{equation}\label{onedimprojpos}
\left(\identitymapn{\Hspdimension}\otimes\Phi\right)
\left(q\otimes\identitymapn{d}\right)\proj{\psi}\left(q\otimes\identitymapn{d}\right)
\geqslant 0\,\forall_{\psi\in\HilbertSp\otimes\HilbertSp}.
\end{equation}
This is the same as
\begin{equation}
\label{onedimprojpostwo}
 \innerpr{\phi}{\left(\identitymapn{\Hspdimension}\otimes\Phi\right)
\left(q\otimes\identitymapn{d}\right)\proj{\psi}\left(q\otimes\identitymapn{d}\right)
\phi}\geqslant 0\,\forall_{\psi,\phi\in\HilbertSp\otimes\HilbertSp}.
\end{equation}
Let $\psi=\sum_{i,j=1}^{\Hspdimension}\psi^{ij}f_i\otimes e_j$ and 
$\phi=\sum_{i,j=1}^{\Hspdimension}\phi^{ij}f_i\otimes e_j$. 
Because of the assumed form of $q$, in index notation the condition 
\eqref{onedimprojpostwo} reads
\begin{equation}\label{onedimprojindx}
\sum_{r,s=1}^{\Hspdimension}\sum_{j,m=1}^{\Hspdimension}\sum_{i,l=1}^k
\cconjb{\phi^{ir}}\Phi_{rs,jm}\psi^{ij}\cconjb{\psi^{lm}}\phi^{ls}\geqslant 0
\end{equation}
for all $\seq{\psi^{ij}}{i,j=1}{i=k,j=\Hspdimension},
\seq{\phi^{lm}}{l,m=1}{l=k,m=\Hspdimension}\subset\setC$. 
Here $\Phi_{rs,jm}$ denote the matrix elements of $\Phi$ with respect to the 
standard basis of $\linearopsb{\HilbertSp}$, 
$\Phi\left(e_{jm}\right)=\sum_{r,s=1}^{\Hspdimension}\Phi_{rs,jm}e_{rs}$.
But eq. \eqref{onedimprojindx} is the same as
\begin{equation}\label{kposwithinnerpr}
 \innerpr{\phi}{\left(\left(\identitymapn{k}\otimes\Phi\right)\proj{\psi}\right)\phi}\geqslant 0\,
\forall_{\psi,\phi\in\setC^k\otimes\HilbertSp}.
\end{equation}
This condition means that $\left(\identitymapn{k}\otimes\Phi\right)\proj{\psi}
\geqslant 0$ 
for any one-dimensional projector $\proj{\psi}$ on $\setC^k\otimes\HilbertSp$, 
which is equivalent to $k$-positivity of $\Phi$.
\end{proof}
\end{lemma}

If $\Phi$ is $k$-positive for every $k\in\setN$, we call it 
\textit{completely positive}. 
We shall denote the family of completely 
positive maps with $\cpositivitymapsb{\HilbertSp}$. Obviously, 
$\cpositivitymapsb{\HilbertSp}=\bigcap_{k\in\setN}\kpositivitymapsb{k}{\HilbertSp}$,
 but it is also a well known fact \cite{Cho75a} that for $k\geqslant d$,
 we get $\kpositivitymapsb{k}{\HilbertSp}=\cpositivitymapsb{\HilbertSp}$. 
A natural question arises whether the sets $\kpositivitymapsb{k}{\HilbertSp}$ with $k\leqslant d$ are 
all distinct one from another. An affirmative answer can be found in \cite{ChruscinskiKossakowskiSpectral}. 
For $k=1,\ldots,d$, the map
\begin{equation}
  \label{exmplkpositive}
  \phi_{\lambda}:\linearopsb{\HilbertSp}\ni a\mapsto\Tra{a}
   \identitymapn{\Hspdimension}-\frac{\lambda}{d}a
\end{equation}
turns out to be $k$-positive iff $\lambda\geqslant\frac{1}{k}$. 
This is a generalization of the famous example by Choi \cite{Ch72} of a map 
that is $d-1$-positive, but not completely positive,
\begin{equation}
  \label{exmplChoikpositive}
  \phi_{Choi}:\linearopsb{\HilbertSp}\ni a\mapsto\Tra{a}
  \identitymapn{\Hspdimension}-\frac{d}{d-1}a.
\end{equation} 

Consider an operator $a\in\linearopsb{\HilbertSp}$. 
It defines a \textit{similarity map}  (also called \textit{adjoint}): 
$\Ad_a:\linearopsb{\HilbertSp}\ni x\mapsto \hconj{a}xa\in\linearopsb{\HilbertSp}$. 
For any operator $a$ such a map is completely positive. 
As observed by Kraus \cite{Kr71}, any completely positive map can be written 
in the form $\sum_{i=1}^n\Ad_{a_i}$ for some 
$\seq{a_i}{i=1}{n}\subset\linearopsb{\HilbertSp}$ ($n\in\setN$). 
The converse holds trivially, so we get 
$\cpositivitymapsb{\HilbertSp}=\convhull{\left\{\Ad_a|a\in\linearopsb{\HilbertSp}\right\}}$. 
If we impose additional conditions on the operators $a_i$,
we get even stronger properties of $\Phi=\sum_{i=1}^n\Ad_{a_i}$ than 
complete positivity. 

For $k\in\setN$, we say that $\Phi$ is \textit{$k$-superpositive} 
iff $\rk a_i\leqslant k$ for all $i=1,\ldots, n$ ($\rk a_i$ denotes the rank of $a_i$).
We denote the set of $k$-superpositive maps by $\kspositivitymapsb{k}{\HilbertSp}$.
Obviously, $\kspositivitymapsb{k}{\HilbertSp}=\cpositivitymapsb{\HilbertSp}$ for 
$k\geqslant\Hspdimension$. 
It is natural to ask whether the classes $\kspositivitymapsb{k}{\HilbertSp}$ with 
$k\leqslant\Hspdimension$ 
are all distinct one from another.
 It turns out that they are, as follows from the Proposition \ref{propkSPdistinct} 
at the end of this section. 
Maps which are $1$-superpositive are simply called \textit{superpositive} \cite{An04}
and we abbreviate the notation $\kspositivitymapsb{1}{\HilbertSp}$ to 
$\spositivitymapsb{\HilbertSp}$.

All the sets of operators that we introduced above have their corresponding left 
transposed partners. 
For any ${\mathcal A}\subset\linearmapsb{\HilbertSp}$, we define
\begin{equation}
\label{cosetdef}
 \coset{\mathcal A}:=\left\{\transpos\circ \Phi|\Phi\in{\mathcal A}\right\},
\end{equation}
where $\transpos$ is the transpose map.
It is customary that the name of $\coset{\mathcal A}$ differs from the 
name of $\mathcal A$ by a ``co'' suffix. For example, 
$\coset{\cpositivitymapsb{\HilbertSp}}$ is called the set of 
\textit{completely copositive} maps. One can easily check that 
$\positivitymapsb{\HilbertSp}=\coset{\positivitymapsb{\HilbertSp}}$ and
$\spositivitymapsb{\HilbertSp}=\coset{\spositivitymapsb{\HilbertSp}}$.

As a conclusion of the above discussion, we get the following chain of inclusions
\begin{equation}
\label{chainincl1}
  \kspositivitymapsb{\!\!}{\!\HilbertSp}\subsetneq 
\kspositivitymapsb{\!2}{\!\HilbertSp}
  \subsetneq  
\!\ldots\!  \subsetneq\kspositivitymapsb{\!d-1}{\!\HilbertSp}
  \subsetneq
\cpositivitymapsb{\!\HilbertSp}
  \subsetneq
\kpositivitymapsb{\!d-1}{\!\HilbertSp}
  \subsetneq  
\!\ldots\! \subsetneq\kpositivitymapsb{\!2}{\!\HilbertSp}
  \subsetneq
\kpositivitymapsb{\!\!}{\!\HilbertSp},
\end{equation}
see columns b) and a) in Table 1.
Finally, we define the following three families of maps ($k,m\in\setN$),
\begin{eqnarray}
 \kmdecmapsb{k}{m}{\HilbertSp}&:=&\kpositivitymapsb{k}{\HilbertSp}
\lor\cosetb{\kpositivitymapsb{m}{\HilbertSp}},\label{kmdecdef}\\
 \kmpositivitymapsb{k}{m}{\HilbertSp}&:=&\kpositivitymapsb{k}{\HilbertSp}
\cap\cosetb{\kpositivitymapsb{m}{\HilbertSp}},\label{kmposdef}\\
 \kmspositivitymapsb{k}{m}{\HilbertSp}&:=&\kspositivitymapsb{k}{\HilbertSp}
\cap\cosetb{\kspositivitymapsb{m}{\HilbertSp}}\label{kmsepdef}.
\end{eqnarray}
We call them \textit{$\left(k,m\right)$-decomposable}, \textit{$\left(k,m\right)$-positive} 
and \textit{$\left(k,m\right)$-superpositive} maps, respectively. Obviously, 
$\kmpositivitymapsb{k}{0}{\HilbertSp}=\kpositivitymapsb{k}{\HilbertSp}$, 
$\kmspositivitymapsb{k}{0}{\HilbertSp}=\kspositivitymapsb{k}{\HilbertSp}$, 
$\kmpositivitymapsb{0}{m}{\HilbertSp}=\cosetb{\kpositivitymapsb{m}{\HilbertSp}}$ 
and $\kmspositivitymapsb{0}{m}{\HilbertSp}=\cosetb{\kspositivitymapsb{m}{\HilbertSp}}$, 
so all the previously discussed classes of maps are included in the definitions \eqref{kmposdef} and \eqref{kmsepdef}. 
It is also easy to see that $\coset{\kmdecmapsb{k}{m}{\HilbertSp}}=\kmdecmapsb{m}{k}{\HilbertSp}$, 
$\coset{\kmpositivitymapsb{k}{m}{\HilbertSp}}=\kmpositivitymapsb{m}{k}{\HilbertSp}$
 and $\coset{\kmspositivitymapsb{k}{m}{\HilbertSp}}=
\kmspositivitymapsb{m}{k}{\HilbertSp} $ in general. 
Note that similar families of  maps and inclusion relations between 
them were analyzed by Chru{\'s}ci{\'n}ski and Kossakowski \cite{CK07},
who called $k$-superpositive maps \textit{partially entanglement breaking channels}. 
In \cite{Cl05} the author defines a family of maps which he calls ``$2$-decomposable'', but they correspond to $\kmspositivitymapsb{0}{2}{\HilbertSp}$ in our notation. That is, we call them ``$2$-supercopositive maps''. On the other hand, the families $\kmdecmapsb{2}{2}{\setC^3}$ and $\kmdecmapsb{2}{2}{\setC^4}$, which we would call $2$-decomposable, appeared many times in the context of \textit{atomic maps} \cite{Cho80,TT88,Ha98}. An element of $\linearmapsb{\HilbertSp}$ is called atomic iff it does not belong to $\kmdecmapsb{2}{2}{\HilbertSp}$. In particular, in \cite{Ha98} it was proved that all the known generalized indecomposable Choi maps of $\linearopsb{\setC^3}$ are atomic. This falsifies the possible conjecture that the St\o rmer-Woronowicz theorem (\!\cite{St63}, \cite{Wo76}) has a generalization of the form $\positivitymapsb{\setC^n}=\kmdecmapsb{n-1}{n-1}{\setC^n}$.

Linear operators on $\linearopsb{\HilbertSp}$ (``maps'') can be 
identified with corresponding elements of 
$\linearopsb{\HilbertSp\otimes\HilbertSp}$ (``operators''). 
In the following, we shall introduce the $\linearopsb{\HilbertSp\otimes\HilbertSp}$
 counterparts of the families of maps that we defined above.

Let $\Phi$ be an element of $\linearmapsb{\HilbertSp}$. Following 
Jamio{\l}kowski \cite{Ja72} and Choi \cite{Cho75a}, we define
\begin{equation}\label{Choimatrix}
 \Choimatr{\Phi}:=\sum_{i,j=1}^{\Hspdimension}e_{ij}
\otimes\Phi\left(e_{ij}\right)=\left(\identitymap\otimes\Phi\right)\proj{\Psi_+},
\end{equation}
where $\Psi_+=\sum_{i}e_i\otimes e_i$ is a maximally entangled state on 
$\HilbertSp\otimes\HilbertSp$. 
We shall denote the map $\Phi\mapsto\Choimatr{\Phi}$ by $\Jamiosymb$,
\begin{equation}\label{Jamiolkowskimap}
J:\linearmapsb{\HilbertSp}\ni\Phi\longmapsto\left(\identitymap\otimes\Phi\right)
\proj{\Psi_+}\in\linearopsb{\HilbertSp\otimes\HilbertSp}.
\end{equation}
It is well known \cite{dePillis,Ja72} that 
$\Jamiosymb|_{\hermicitymapsb{\HilbertSp}}$ is an isomorphism between 
$\hermicitymapsb{\HilbertSp}$ and the set of Hermitian operators on 
$\HilbertSp\otimes\HilbertSp$, $\hermitianopsb{\HilbertSp\otimes\HilbertSp}$. 
Since $\positivitymapsb{\HilbertSp}\subset\hermicitymapsb{\HilbertSp}$, 
we shall concentrate on $\Phi|_{\hermicitymapsb{\HilbertSp}}$ in most of what 
follows and we omit the subscript $|_{\hermicitymapsb{\HilbertSp}}$. 
Thus $\Jamiosymb$ can be regarded as a $\setR$-linear isomorphism between the 
$\setR$-linear spaces $\hermicitymapsb{\HilbertSp}$ and 
$\hermitianopsb{\HilbertSp\otimes\HilbertSp}$.

Let us introduce the so-called set of \textit{$k$-block positive operators} 
($k\in\setN$),
\begin{equation}\label{kblockposdef}
 \kbposopsb{k}{\HilbertSp\otimes\HilbertSp}:=
\left\{a\,\vline\innerpr{\sum_{i=1}^k\phi_i\otimes \psi_i}
{a\sum_{l=1}^k\phi_l\otimes \psi_l}\geqslant 0\,
\forall_{\seq{\psi_i}{i=1}{k},
\seq{\phi_i}{l=1}{k}\subset\HilbertSp}\right\},
\end{equation}
where the $a$'s are elements of $\linearopsb{\HilbertSp\otimes\HilbertSp}$. 
We write $\bposopsb{\HilbertSp\otimes\HilbertSp}$ instead of 
$\kbposopsb{1}{\HilbertSp\otimes\HilbertSp}$ and simply call 
$1$-block positive operators 
\textit{block positive}. One can easily prove that 
$\kbposopsb{k}{\HilbertSp\otimes\HilbertSp}\subset
\hermitianopsb{\HilbertSp\otimes\HilbertSp}$
 for arbitrary $k\geqslant 1$ (cf. \cite{MastersThesis}). 
Moreover, we have the following
\begin{proposition}
\label{lemmakbpkpos}{\rm (Generalized Jamio{\l}kowski-Choi theorem)}
 Let $k$ be a positive integer. The sets $\kpositivitymapsb{k}{\HilbertSp}$
 and $\kbposopsb{k}{\HilbertSp\otimes\HilbertSp}$ are isomorphic. We have
\begin{equation}\label{jamisokposkbp}
 \Jamiso{\kpositivitymapsb{k}{\HilbertSp}}=
\kbposopsb{k}{\HilbertSp\otimes\HilbertSp},
\end{equation}
where the isomorphism $\Jamiosymb$ was defined in \eqref{Jamiolkowskimap}.
\begin{proof}Let $\Phi$ be an element of $\hermicitymapsb{\HilbertSp}$. 
We shall prove that $\Phi\in\kpositivitymapsb{k}{\HilbertSp}$ is 
equivalent to $C_{\Phi}\in\kbposopsb{k}{\HilbertSp\otimes\HilbertSp}$ and 
thus we will have proved \eqref{jamisokposkbp}. We start from the following lemma,
\begin{lemma}\label{lemmareshuffling}
 Let $\Phi\in\hermicitymapsb{\HilbertSp}$ and denote by $\Phi_{ij,kl}$ 
the matrix elements of $\Phi$ with respect to the standard basis of 
$\linearopsb{\HilbertSp}$, $\Phi\left(e_{kl}\right)=
\sum_{i,j=1}^{\Hspdimension}\Phi_{ij,kl}e_{ij}$. Let 
$\Choimatr{\Phi}=\left(\Choimatr{\Phi}\right)_{rs,tu}e_{rt}\otimes e_{su}$, 
so that $\left(\Choimatr{\Phi}\right)_{rs,tu}$ are the coefficients of 
$\Choimatr{\Phi}$ with respect to the basis 
$\seq{e_{rt}\otimes e_{su}}{r,t,s,u=1}{\Hspdimension}$. Then we have
\begin{equation}
\label{reshufflingeqn}
 \left(\Choimatr{\Phi}\right)_{ij,kl}=\Phi_{jl,ik}.
\end{equation}
\begin{proof}
By definition (see \eqref{Choimatrix}), 
$\Choimatr{\Phi}=\sum_{r,s=1}^{\Hspdimension}e_{rs}\otimes\Phi\left(e_{rs}\right)$. 
In index notation,
\begin{equation}\label{Choimatrixindexnot}
 \left(\Choimatr{\Phi}\right)_{ij,kl}=
\sum_{r,s=1}^{\Hspdimension}
\left(e_{rs}\otimes\Phi\left(e_{rs}\right)\right)_{ij,kl}=
\sum_{r,s=1}^{\Hspdimension}\left(e_{rs}\right)_{ik}
\left(\Phi\left(e_{rs}\right)\right)_{jl}.
\end{equation}
From \eqref{Choimatrixindexnot} we readily get
\begin{equation}\label{Choimatrixindexnot2}
 \left(\Choimatr{\Phi}\right)_{ij,kl}=\sum_{r,s=1}^{\Hspdimension}
\delta_{ri}\delta_{sk}\left(\Phi\left(e_{rs}\right)\right)_{jl}=
\sum_{r,s=1}^{\Hspdimension}\delta_{ri}\delta_{sk}\Phi_{jl,rs}=\Phi_{jl,ik},
\end{equation}
which is the expected formula. Such a reordering of elements of 
the superoperator $\Phi$,
 first used by Sudarshan et al. \cite{SMR61} to obtain the matrix $\Choimatr{\Phi}$, 
was later called \textit{reshuffling} \cite{BZ06}.
\end{proof}
\end{lemma}

Now we can prove Proposition \ref{lemmakbpkpos}. When applied 
to $\Choimatr{\Phi}$, the $k$-block positivity condition that appears 
in \eqref{kblockposdef} may be rewritten in index notation as
\begin{equation}\label{indexnotkbpcond}
\sum_{r,s=1}^{\Hspdimension}\sum_{j,m=1}^{\Hspdimension}
\sum_{i,l=1}^k\cconjb{\psi_i^{r}}\phi_i^{j}\left(\Choimatr{\Phi}\right)_{rj,sm}
\cconjb{\phi_l^{m}}\psi_l^{s}\geqslant 0 
\end{equation}
for all $\seq{\psi_i^j}{i,j=1}{i=k,j=\Hspdimension},
\seq{\phi_l^m}{l,m=1}{l=k,m=\Hspdimension}\subset\setC$. 
Since this should hold for arbitrary sets of complex numbers 
$\psi_i^j$, $\phi_l^m$, we can complex conjugate all of them in 
\eqref{indexnotkbpcond}. We also change the names of indices like 
$j\leftrightarrow r$ and $m\leftrightarrow s$. After all these changes we get
 as equivalent to \eqref{indexnotkbpcond},
\begin{equation}
\label{indexnotkbpcond3}
 \sum_{r,s=1}^{\Hspdimension}\sum_{j,m=1}^{\Hspdimension}
\sum_{i,l=1}^k\psi_i^{j}\cconjb{\phi_i^{r}}
\left(\Choimatr{\Phi}\right)_{jr,ms}\phi_l^{s}\cconjb{\psi_l^{m}}\geqslant 0, 
\end{equation}
which should hold for all $\seq{\psi_i^j}{i,j=1}{i=k,j=\Hspdimension},
\seq{\phi_l^m}{l,m=1}{l=k,m=\Hspdimension}\subset\setC$.

Using Lemma \ref{lemmareshuffling}, we may rewrite \eqref{indexnotkbpcond3} as
\begin{equation}\label{indexnotkbpcond2}
 \sum_{r,s=1}^{\Hspdimension}\sum_{j,m=1}^{\Hspdimension}\sum_{i,l=1}^k
\cconjb{\psi_i^{r}}\phi_i^{j}\Phi_{rs,jm}\cconjb{\phi_l^{m}}\psi_l^{s}\geqslant 0.
\end{equation}
After small rearrangements, this is precisely condition \eqref{onedimprojindx}. 
The only difference is that the position of the first index in $\phi^{ij}$ and
 in $\psi^{lm}$ was changed, which is not significant.
 As we mentioned in the proof of Lemma \ref{lemmakposchar}, \eqref{onedimprojindx}
 is equivalent to $k$-positivity of $\Phi$ and so is \eqref{indexnotkbpcond2}.
\end{proof}
\end{proposition}

\noindent Proposition \ref{lemmakbpkpos} appears in the early work by Takasaki and Tomiyama, \cite{TT83} (it was also proved in \cite{RA07} using different methods). Thus we have found the $\linearopsb{\HilbertSp\otimes\HilbertSp}$ counterparts 
of the sets $\kpositivitymapsb{k}{\HilbertSp}$. 
In particular, the case $k=1$ gives the relation between positive maps 
and block positive operators, analyzed by Jamio{\l}kowski \cite{Ja72}.
On the other hand, for any $k\geqslant d$ one has that
$\kbposopsb{k} {\HilbertSp\otimes\HilbertSp}=\posopsb{\HilbertSp\otimes\HilbertSp}$.
A similar equality holds between $\kpositivitymapsb{k}{\HilbertSp}$ and 
$\cpositivitymapsb{\HilbertSp}$ for $k\geqslant\Hspdimension$. 
Using Proposition \ref{lemmakbpkpos}, we recover the Choi's well known result \cite{Cho75a},
\begin{proposition}[Choi]
 \label{Choithm}
 The set of completely positive maps of $\linearopsb{\HilbertSp}$ is isomorphic to 
the set of positive operators on composed Hilbert space, 
\begin{equation}\label{Choithmeq}
 \Jamiso{\cpositivitymapsb{\HilbertSp}}=\posopsb{\HilbertSp\otimes\HilbertSp}.
\end{equation}
\qed
\end{proposition}
Thus for intermediate integer values, $k=2,\dots,d-1$,
we get a kind of discrete interpolation between the theorems of Jamio{\l}kowski 
and Choi.

\medskip

To find the sets of operators corresponding to $k$-superpositive maps, 
we shall need the following lemma,

\begin{lemma}
\label{lemmaAdprojectors}
     Let $a\in\linearopsb{\HilbertSp}$. Then
\begin{equation}
\label{imageofAda}
 \Choimatr{\Ad_a}=\proj{\alpha},
\end{equation}
where $\alpha\in\HilbertSp\otimes\HilbertSp$, $r:=\rk a$ and
\begin{equation}
\label{omegavec}
 \alpha=\sum_{l=1}^r \phi_l\otimes\psi_l
\end{equation}
for some orthogonal vectors $\seq{\phi_i}{i=1}{r}, 
\seq{\psi_j}{j=1}{r}\subset\HilbertSp$. 
Any operator $\proj{\alpha}$ with $\alpha$ of the form \eqref{omegavec} 
can be obtained as 
$\Choimatr{\Ad_{a}}$ for some $a\in\linearopsb{\HilbertSp}$.
\begin{proof}
 From the polar decomposition of $a$, we have 
$a=\sum_{l=1}^{r}\sqrt{\lambda_l}U\proj{\psi_l}$, where the $\lambda_l$'s 
are the eigenvalues of $\left|a\right|:=\sqrt{\hconj{a}a}$, 
$U$ is a unitary operator on $\HilbertSp$
 and the vectors $\psi_l\in\HilbertSp$ are orthonormal. 
By the definition \eqref{Choimatrix},
\begin{equation}
\label{ChoimatrixAd}
\Choimatr{\Ad a}=\sum_{l,m=1}^{r}
\sum_{i,j=1}^{\Hspdimension}e_{ij}\otimes\sqrt{\lambda_l\lambda_m}
\innerpr{\psi_l}{\hconj{U}e_{ij}
U\psi_m}\diad{\psi_l}{\psi_m}
\end{equation}
Define $\tilde\psi_l=\sqrt{\lambda_l}
\sum_{i=1}^{\Hspdimension}\sum_{j=1}^{\Hspdimension}U^i_j\psi_l^je_i$ and $\phi_l=\sqrt{\lambda_l}\sum_{i=1}^{\Hspdimension}
\cconjb{\sum_{j=1}^{\Hspdimension}U^i_j\psi_l^j}e_i$, 
where $\psi_l=\sum_{j=1}^{\Hspdimension}\psi_l^je_j$ and 
$U^i_j$ are matrix elements of $U$. The vectors $\phi_i$ are mutually orthogonal. We get
\begin{equation}
\label{ChoimatrixAd2}
 \Choimatr{\Ad a}=\sum_{l,m=1}^{r}\sum_{i,j=1}^{\Hspdimension}
\innerpr{\tilde\psi_l}{e_{ij}\tilde\psi_m}e_{ij}\otimes\diad{\psi_l}{\psi_m}
\end{equation}
It is easy to show that $\sum_{i,j=1}^{\Hspdimension}\innerpr{\tilde\psi_l}{e_{ij}
\tilde\psi_m}e_{ij}=\diad{\phi_l}{\phi_m}$. 
Hence \eqref{ChoimatrixAd2} can be rewritten as
\begin{equation}
\label{ChoimatrixAd3}
 \Choimatr{\Ad a}=\sum_{l,m=1}^{r}\diad{\phi_l}{\phi_m}\otimes\diad{\psi_l}{\psi_m},
\end{equation}
which equals $\proj{\alpha}$ for $\alpha=\sum_{l=1}^{r}\phi_l\otimes\psi_l$. 
This proves the main part of the lemma. The fact that any projector 
$\proj{\alpha}$ can be
 obtained in this way follows from the calculation of $\Choimatr{\Ad_a}$ 
for $a=\sum_{i=1}^k\diad{\tilde\phi_i}{\psi_i}$.
\end{proof}
\end{lemma}
Using Lemma \ref{lemmaAdprojectors}, we can prove the promised 
result that all the  sets $\kpositivitymapsb{k}{\HilbertSp}$ for 
$k=1,\ldots,\HilbertSp$ 
are distinct. We have the following


\begin{proposition}
\label{propkSPdistinct}
  Let $k\leqslant\Hspdimension$ be a positive integer. 
  Let $a\in\linearopsb{\HilbertSp}$ and $\rk a=k$. 
The similarity map $\Ad_{a}$ is an element of $\kspositivitymapsb{k}{\HilbertSp}$, 
but not of $\kspositivitymapsb{k-1}{\HilbertSp}$.

\begin{proof}
Let $a$ be as in the assumptions of the proposition. 
Obviously, $\Ad_a$ is an element of $\kspositivitymapsb{k}{\HilbertSp}$. 
Let us assume $\Ad_a=\sum_i\Ad_{a_i}$ for some nonzero operators $\seq{a_i}{i=1}{m}\subset
\linearopsb{\HilbertSp}$. 
By calculating the Choi matrices of both sides 
of this equality, we get from Lemma \ref{lemmaAdprojectors}
\begin{equation}
\label{projectorssumeq} 
 \proj{\alpha}=\sum_{l=1}^m\proj{\alpha_l}
\end{equation}
for some $m\in\setN$ and nonzero vectors $\alpha\in\HilbertSp$, 
$\seq{\alpha_l}{l=1}{m}\subset\HilbertSp$ such that $\Choimatr{a}=\proj{\alpha}$ 
and $\Choimatr{a_l}=\proj{\alpha_l}$. But \eqref{projectorssumeq} can only hold if 
all the vectors $\alpha_l$ are scalar multiples of $\alpha$. According to 
Lemma \ref{lemmaAdprojectors},  $\alpha$ is of the form 
$\sum_{l=1}^k\phi_l\otimes\psi_l$, so all the vectors $\alpha_l$ have to be of 
the same form as well. Using Lemma \ref{lemmaAdprojectors} again, we conclude that
 $\rk a_l=k$. Since we made no assumptions about the $a_l$'s, 
the equality $\rk a_l=k$ implies that $\Ad_a$ cannot be an element of 
$\kspositivitymapsb{k-1}{\HilbertSp}$. This proves our assertion\footnote{A  
simpler proof of Proposition \ref{propkSPdistinct} can be obtained by noting that the
Choi matrix $\Choimatr{\Ad_a}$ is a positive rank one operator, and so are all the  
Choi matrices $\Choimatr{\Ad_{a_i}}$, hence the $\Ad_{a_i}$ are  scalar multiples of  
$\Ad_a$.  We have kept the longer proof because of its 
connection  
with Lemma \ref{lemmaAdprojectors}}.
\end{proof}
\end{proposition}
In short, Proposition \ref{propkSPdistinct} 
implies that $\kspositivitymapsb{k-1}{\HilbertSp}\subsetneq
\kspositivitymapsb{k}{\HilbertSp}$ 
for $k\leqslant\Hspdimension$, as we already mentioned above.

Lemma \ref{lemmaAdprojectors} can as well be used to find the families 
of operators in 
$\linearopsb{\HilbertSp\otimes\HilbertSp}$ corresponding to $k$-superpositive maps.
 By the very definition of $\kspositivitymapsb{k}{\HilbertSp}$,
 an element  $\Phi\in\linearmapsb{\HilbertSp}$ is $k$-superpositive 
iff it is of the form 
$\sum_{l=1}^m\Ad_{a_l}$ for some $m\in\setN$ and 
$\seq{a_l}{l=1}{m}\subset\linearopsb{\HilbertSp}$ 
such that $\rk a_l\leqslant k$ for all $l=1,\ldots,m$.
According to Lemma \ref{lemmaAdprojectors}, this is the same as
\begin{equation}
  \label{Adsumksep}
 \Choimatr{\Phi}=\sum_{i,j=1}^k\sum_{l=1}^m\diad{\phi^{\left(l\right)}_i
\otimes\psi^{\left(l\right)}_i}{\phi^{\left(l\right)}_j
\otimes\psi^{\left(l\right)}_j}
\end{equation}
for some $m\in\setN$ and sets of vectors\footnote{We do not assume the vectors 
to be nonzero}
 $\seq{\phi^{\left(l\right)}_i}{i=1}{k}, \seq{\psi^{\left(l\right)}_j}{j=1}{k}
\in\HilbertSp$, where $l=1,\ldots,m$. 
Obviously, operators on the right hand side of \eqref{Adsumksep} make up 
the convex cone spanned by the 
positive rank $1$ operators $\sum_{i,j=1}^k\diad{\phi_i\otimes\psi_i}{\phi_j\otimes\psi_j}$. 
This is nothing else as the definition of an operator
with the \textit{Schmidt number} equal to $k$  - see  \cite{TH00,SBL01,RA07}.

Thus we get the following
\begin{proposition}
\label{propkSPkSep}
 Let $k$ be a positive integer.
 Let us define the set of \textit{$k$-separable operators} on 
$\HilbertSp\otimes\HilbertSp$
(equivalent to the set of operators with Schmidt number less than or equal to $k$),  
\begin{equation}\label{ksepdef}
 \ksepopsb{k}{\HilbertSp\otimes\HilbertSp}:=
\convhull{\left\{\sum_{i,j=1}^k\diad{\phi_i\otimes\psi_i}{\phi_j\otimes\psi_j}\,
\vline\seq{\phi_i}{i=1}{k},\seq{\psi_j}{j=1}{k}\subset\HilbertSp\right\}}.
\end{equation}
Thus the set of $k$-superpositive maps is isomorphic to 
$\ksepopsb{k}{\HilbertSp\otimes\HilbertSp}$,
\begin{equation}
\label{ksepisokSP}
 \Jamiso{\kspositivitymapsb{k}{\HilbertSp}}=\ksepopsb{k}{\HilbertSp\otimes\HilbertSp}.
\end{equation}
\qed
\end{proposition}

\noindent We can now write a chain of inclusions corresponding 
to \eqref{chainincl1},
\begin{equation}\label{chainincl2}
\sepops\subsetneq\ldots
\subsetneq\ksepops{\left(d-1\right)}
\subsetneq\posops
\subsetneq\kbposops{\left(d-1\right)}
\subsetneq\ldots\subsetneq\bposops
\end{equation}
(we omit the brackets $\left(\HilbertSp\otimes\HilbertSp\right)$ 
to fit the formula into the page and write $\sepops$ instead of
 $\ksepops{1}$ to simplify notation. 
The elements of $\sepopsb{\HilbertSp\otimes\HilbertSp}$ are called
 \textit{separable} operators).
This chain of inclusions, studied earlier in \cite{CK07}, 
corresponds to columns b') and a') in Table \ref{table1cones} on page \pageref{table1cones}.

To find the sets of operators corresponding to completely 
copositive ($\coset{\cpositivitymapsb{\HilbertSp}}$), 
k-copositive ($\coset{\kpositivitymapsb{k}{\HilbertSp}}$) and 
k-supercopositive maps ($\coset{\kspositivitymapsb{k}{\HilbertSp}}$), 
we use the following lemma
\begin{lemma}
\label{lemmapartialtr}
 Let $\mathcal A$ be a subset of $\linearmapsb{\HilbertSp}$ and 
$\Jamiso{\mathcal{A}}\subset\linearopsb{\HilbertSp\otimes\HilbertSp}$. 
We have
\begin{equation}\label{imageJtranspose}
 \Jamiso{\coset{\mathcal{A}}}=\parttrb{\Jamiso{\mathcal{A}}}:=
\left\{\parttrb{a}\,\vline\, a\in\Jamiso{\mathcal{A}}\right\},
\end{equation}
\begin{proof}
 From the definition \eqref{Choimatrix}, we have
\begin{equation}
  \label{Choimatratranspose}
  \Choimatr{\transpos\circ\Phi}=\left(\identitymap\otimes
\left(\transpos\circ\Phi\right)\right)
  \proj{\Psi_+}=\parttrb{}\left(\identitymap\otimes\Phi\right)
\proj{\Psi_+}=\parttrb{}\Choimatr{\Phi}.
\end{equation}
This gives us $\Jamiso{\transpos\circ\Phi}=\parttrb{\Jamiso{\Phi}}$,
 which proves the lemma.
\end{proof}
\end{lemma}
\noindent The map $\parttr$ that appears in Lemma \ref{lemmapartialtr} is called 
\textit{partial transposition}. Using the lemma, we trivially get
\begin{proposition}\label{propcosetsoperators}Let $k$ be a positive integer. 
We have the correspondences
\begin{eqnarray}
 \Jamiso{\coset{\cpositivitymapsb{\HilbertSp}}}&=&
\parttrb{\posopsb{\HilbertSp\otimes\HilbertSp}}\label{coposops},\\
\Jamiso{\coset{\kpositivitymapsb{k}{\HilbertSp}}}&=&
\parttrb{\kbposopsb{k}{\HilbertSp\otimes\HilbertSp}}\label{cokposops},\\
\Jamiso{\coset{\kspositivitymapsb{k}{\HilbertSp}}}&=&
\parttrb{\ksepopsb{k}{\HilbertSp\otimes\HilbertSp}}\label{coksepops}.
\end{eqnarray}
\qed
\end{proposition}
\noindent The sets $\kmdecmapsb{k}{m}{\HilbertSp}$, 
$\kmpositivitymapsb{k}{m}{\HilbertSp}$ and $\kmspositivitymapsb{k}{m}{\HilbertSp}$ 
also have their $\linearopsb{\HilbertSp\otimes\HilbertSp}$ counterparts,
\begin{proposition}\label{kmdecsepposops}Let $k,m$ be positive integers. 
We have
 \begin{eqnarray}
  \Jamiso{\kmdecmapsb{k}{m}{\HilbertSp}}&=&
\kbposopsb{k}{\HilbertSp\otimes\HilbertSp}\lor
\parttrb{\kbposopsb{m}{\HilbertSp\otimes\HilbertSp}},
\label{kmdecopdef}\\ 
\Jamiso{\kmpositivitymapsb{k}{m}{\HilbertSp}}&=&
\kbposopsb{k}{\HilbertSp\otimes\HilbertSp}\cap
\parttrb{\kbposopsb{m}{\HilbertSp\otimes\HilbertSp}},
\label{kmposopdef}\\
 \Jamiso{\kmspositivitymapsb{k}{m}{\HilbertSp}}&=&
\ksepopsb{k}{\HilbertSp\otimes\HilbertSp}\cap
\parttrb{\ksepopsb{m}{\HilbertSp\otimes\HilbertSp}}
\label{kmsepopdef}.
\end{eqnarray}
\qed
\end{proposition}


\section{Relations between $k$-positive and $k$-superpositive maps. Other relations}
\label{sectionRelations}
It is a well known fact that $\hermitianopsb{\HilbertSp\otimes\HilbertSp}$ is 
a $\Hspdimension^4$-dimensional vector space over $\setR$ and it is equipped with 
the symmetric \textit{Hilbert-Schmidt product},
\begin{equation}\label{HSprod}
 \HSprod{a}{b}:=\Trb{\hconj{a}b}=\Trb{ab},
\end{equation}
where $a,b\in\hermitianopsb{\HilbertSp\otimes\HilbertSp}$, and the last equality holds due to the Hermiticity of $a$.

Let $A$ be a cone in $\hermitianopsb{\HilbertSp\otimes\HilbertSp}$. 
We define the \textit{dual cone} of $A$,
\begin{equation}\label{dualconedef}
\dualityops{A}:=\left\{b\in\hermitianopsb{\HilbertSp\otimes\HilbertSp}\,\vline\,
\HSprod{a}{b}\geqslant 0\,\forall_{a\in A}\right\}.
\end{equation}
By comparing the definitions \eqref{kblockposdef} and \eqref{ksepdef}, 
we easily get
\begin{proposition}\label{kBPdualofkP}$\kbposopsb{k}{\HilbertSp\otimes\HilbertSp}=
\dualityopsb{\ksepopsb{k}{\HilbertSp\otimes\HilbertSp}}$
\begin{proof}
 Follows directly from the definition of $\kbposopsb{k}{\HilbertSp\otimes\HilbertSp}$
if we observe that
\begin{equation}\label{relTrInner}
 \innerpr{\sum_{i=1}^k\phi_i\otimes\psi_i}
{a\sum_{j=1}^k\phi_j\otimes\psi_j}=
\Trb{a\sum_{i,j=1}\diad{\phi_j\otimes\psi_j}{\phi_i\otimes\psi_i}}.
\end{equation}
\end{proof}
\end{proposition}
\noindent By substituting $k=\Hspdimension$, we get 
$\dualityopsb{\posopsb{\HilbertSp\otimes\HilbertSp}}=
\posopsb{\HilbertSp\otimes\HilbertSp}$, 
which was discussed  in \cite{ZB04,CK07}, 
and  may easily be proved directly.
Remember that we have
$\ksepopsb{\Hspdimension}{\HilbertSp\otimes\HilbertSp}=
\kbposopsb{\Hspdimension}{\HilbertSp\otimes\HilbertSp}=
\posopsb{\HilbertSp\otimes\HilbertSp}$.

From the existence of separating hyperplanes in $\setR^n$ 
(cf. Theorem 14.1 in \cite{Rockafellar}) it follows that 
$\dualityopsb{\dualityops{A}}=\bar A$ for any cone 
$A\in\hermitianopsb{\HilbertSp\otimes\HilbertSp}$. In particular,
\begin{equation}
\label{bidualthm}
 \dualityopsb{\dualityops{A}}=A
\end{equation}
for a closed cone $A\subset\hermitianopsb{\HilbertSp\otimes\HilbertSp}$. 
We call this fact the \textit{bidual theorem}. As a consequence, we have
\begin{proposition}\label{kSepdualofkSP}
 $\ksepopsb{k}{\HilbertSp\otimes\HilbertSp}=
\dualityopsb{\kbposopsb{k}{\HilbertSp\otimes\HilbertSp}}$
\begin{proof}
 In is easy to show that the set $\ksepopsb{k}{\HilbertSp\otimes\HilbertSp}$ is 
closed (cf. e.g. \cite{MastersThesis}). 
Thus we can use the bidual theorem together with 
Proposition \ref{kBPdualofkP} to prove our assertion.
\end{proof}
\end{proposition}
 
\begin{figure}
\begin{center} 
\includegraphics[scale=0.33]{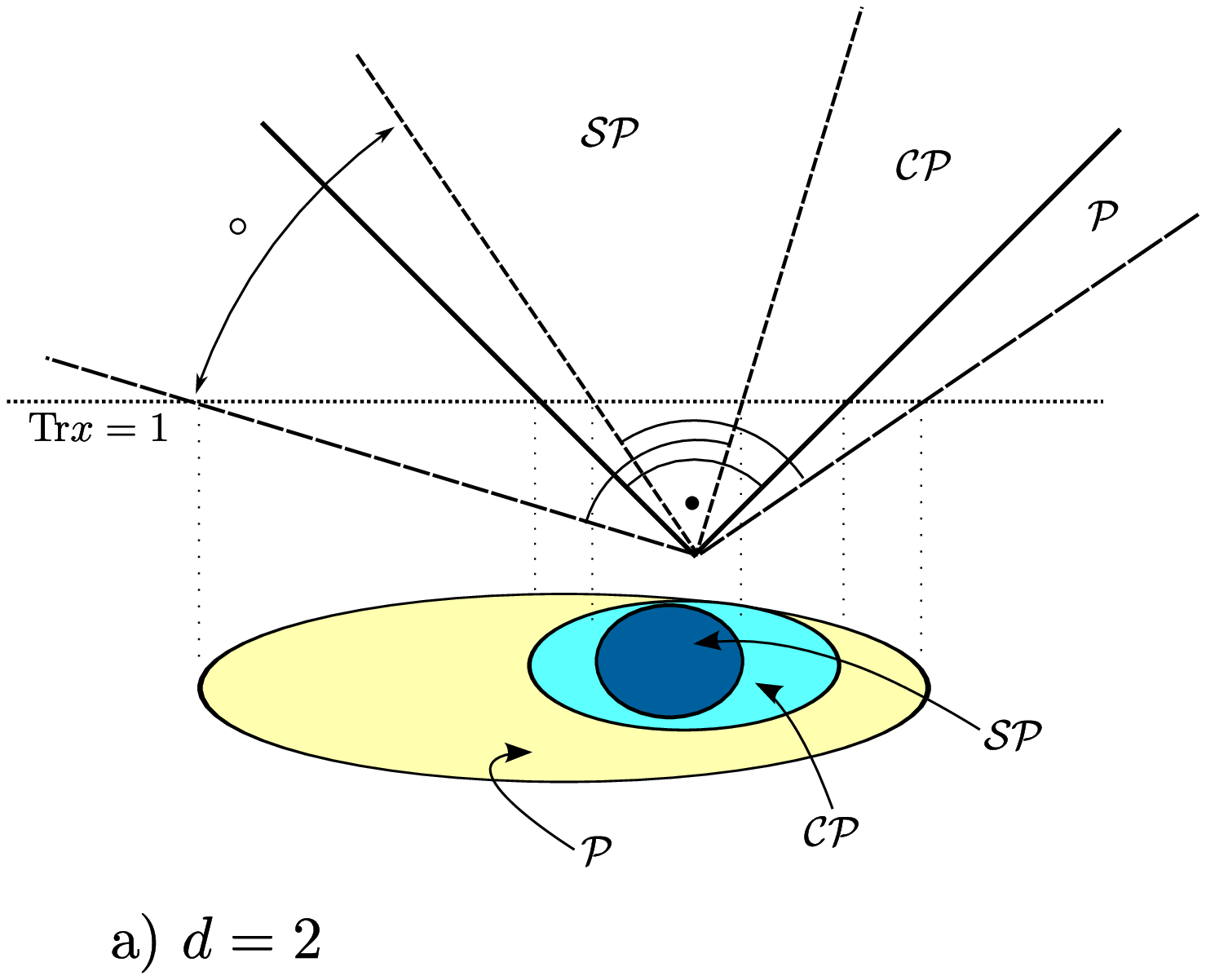}\hskip 3 mm
\includegraphics[scale=0.33]{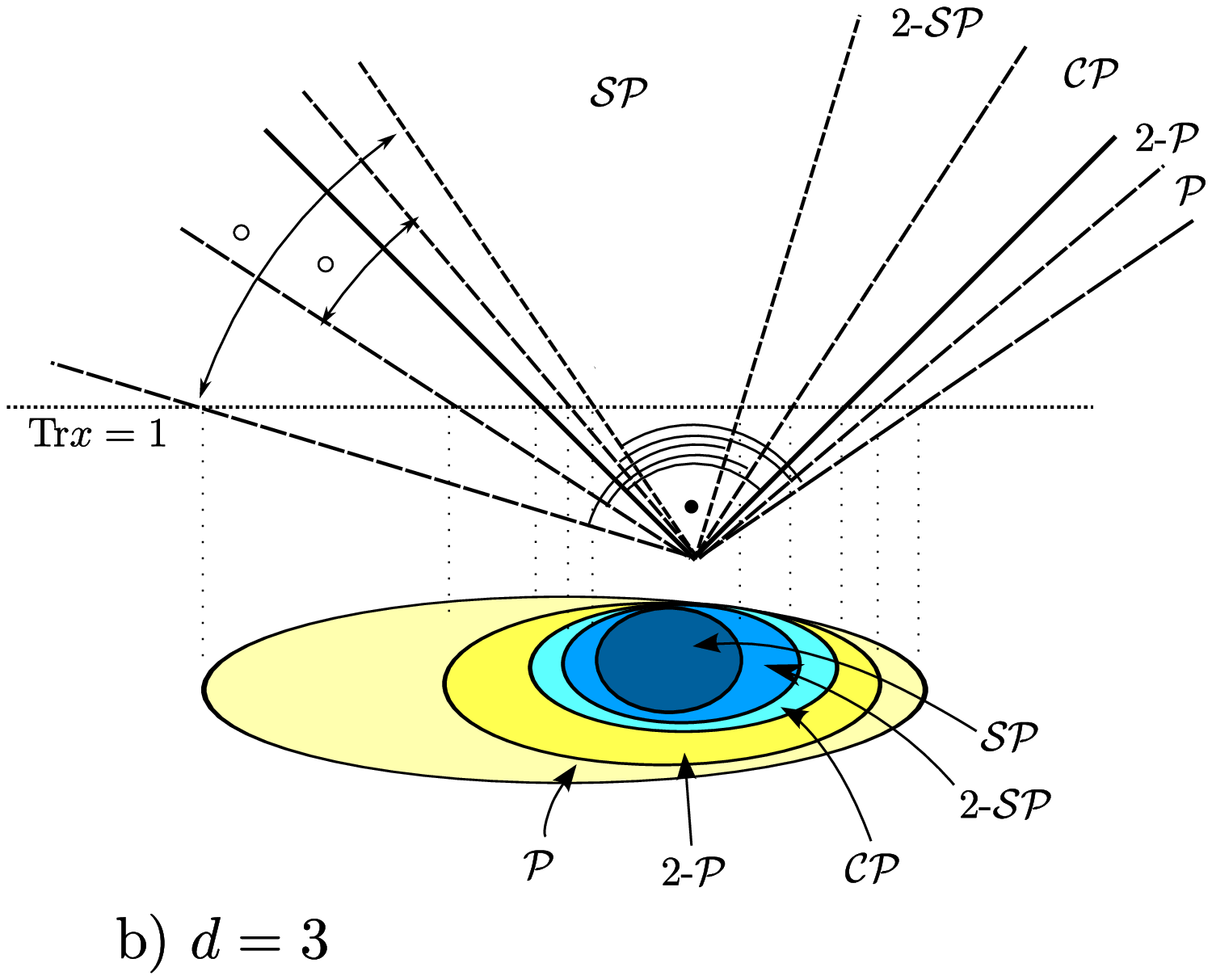}
\caption{\label{figuredualities0}
Cones of positive maps:   a) $d=2$, self-dual cone 
 $\cpositivitymapsb{\HilbertSp}=\dualityopsb{\cpositivitymapsb{\HilbertSp}}$ and
 a pair of dual cones  $\positivitymapsb{\HilbertSp}=\dualityops{\spositivitymapsb{\HilbertSp}}$;
b) case $d=3$ with yet another pair of dual cones
$\kpositivitymapsb{2}{\HilbertSp}=\dualityopsb{\kspositivitymapsb{2}{\HilbertSp}}$.
The plot  above shows unbounded cones and the normalization hyperplane $\Tra{x}=1$, 
while the convex sets below represent their cross-sections. 
The same sketch is applicable to the corresponding cones of
block positive, positive semidefinite and separable operators.
}
\end{center}
\end{figure}

Using the natural duality in $\hermitianopsb{\HilbertSp\otimes\HilbertSp}$, 
we can introduce an analogous operation in $\hermicitymapsb{\HilbertSp}$. 
Let $\mathcal{X}\subset\hermicitymapsb{\HilbertSp}$ be a convex cone. 
We define the \textit{dual cone} of $\mathcal{X}$ as
\begin{equation}
\label{defdualconeinLH}
 \dualitymaps{\mathcal{X}}:=\left\{\Phi\in\hermicitymapsb{\HilbertSp}\,\vline\,
\Trb{C_{\Phi}C_{\Psi}}\geqslant 0\,\forall_{\Psi\in\mathcal{X}}\right\}.
\end{equation}
It is easy to notice that \eqref{defdualconeinLH} can as well be written as
\begin{equation}\label{alterdefdualconeinLH}
 \dualitymapsb{\mathcal{X}}=\Jamiosymb^{-1}\left(\dualityopsb{\Jamiso{\mathcal{X}}}\right),
\end{equation}
which makes the definition \eqref{defdualconeinLH} transparent.
 As a direct consequence of \eqref{alterdefdualconeinLH} and Propositions \ref{lemmakbpkpos} and \ref{kBPdualofkP}, we obtain
\begin{proposition}\label{kposdualksp}
 $\kpositivitymapsb{k}{\HilbertSp}=\dualitymaps{\kspositivitymapsb{k}{\HilbertSp}}$
\qed
\end{proposition}
\noindent In a similar way, using Propositions \ref{propkSPkSep} and 
\ref{kSepdualofkSP}, we obtain
\begin{proposition}\label{ksposdualkpos}
 $\kspositivitymapsb{k}{\HilbertSp}=
\dualitymaps{\kpositivitymapsb{k}{\HilbertSp}}$
\qed
\end{proposition}
\noindent This result was given in a slightly less explicit way in \cite{EK00}.

Remembering that $\kspositivitymapsb{\Hspdimension}{\HilbertSp}
=\kpositivitymapsb{\Hspdimension}{\HilbertSp}=
\cpositivitymapsb{\HilbertSp}$, we easily obtain from Proposition 
\ref{kposdualksp} or \ref{ksposdualkpos} the relation 
$\dualitymaps{\cpositivitymapsb{\HilbertSp}}=\cpositivitymapsb{\HilbertSp}$. 
\emph{The set of completely positive maps is self-dual}.
    
\begin{figure}
\begin{center} 
\includegraphics[scale=0.5]{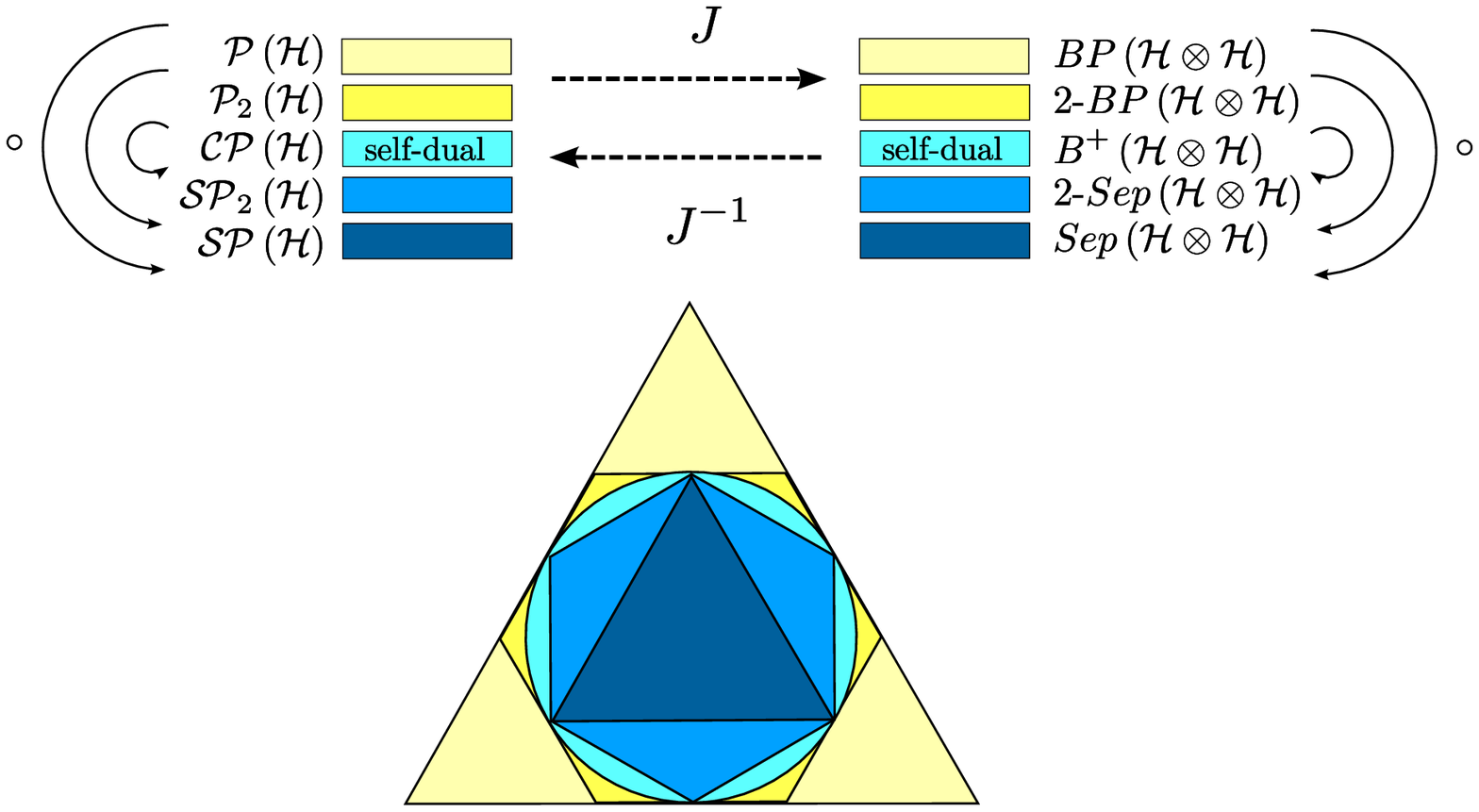}
\caption{\label{figuredualities1}
A schematic picture of the chain of inclusions 
$\spositivitymapsb{\HilbertSp}\subsetneq\kspositivitymapsb{2}{\HilbertSp}
\subsetneq\cpositivitymapsb{\HilbertSp}
\subsetneq\kpositivitymapsb{2}{\HilbertSp}\subsetneq\positivitymapsb{\HilbertSp}$ 
($d\geqslant 3$),
 which takes into account the duality relations expressed 
in Propositions \ref{kposdualksp} and \ref{ksposdualkpos}.
The same sketch represents also the inclusion relations
among  the sets of normalized operators, 
which correspond to sets of maps with respect to the 
Jamio{\l}kowski isomorphism $\Jamiosymb$.}
\end{center}
\end{figure}

The relations expressed in Propositions \ref{kposdualksp} 
and \ref{ksposdualkpos}
 can be depicted as in Figure \ref{figuredualities0},
which shows the the cones of block-positive, positive and 
separable operators for $d=2$ and $d=3$.
Note that the self-dual cone for positive operators is represented by
 the right-angled triangle. 
The same sketch represents also the corresponding cones of maps.
In physical application one is often interested in a set of 
normalized operators.
For instance, the trace normalization $\Tra{x}=1$ corresponds 
to a hyperplane,
represented by a horizontal line.

 The cross-section of such a normalization hyperplane with 
each cone gives bounded convex sets of a finite volume estimated in \cite{SWZ08}.
Their structure for $d=3$ is sketched in  Fig. \ref{figuredualities1}.
 The picture is exact in the sense that there exist convex 
cones in $\setR^3$ such that their section by an appropriately chosen
 plane gives the above sets which  fulfill the duality relations 
in accordance with Propositions \ref{kposdualksp} and \ref{ksposdualkpos}. 
For example, the circle in Figure \ref{figuredualities1} is a section of a cone of 
aperture $\pi/2$ by a plane perpendicular to its axis. 
The cone is self-dual, just as the 
set $\cpositivitymapsb{\HilbertSp}$ which it represents.

\begin{figure}[h]
\begin{center} 
\includegraphics[scale=0.5]{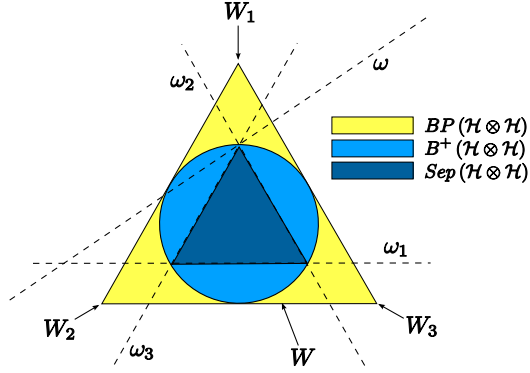}\hskip 2 cm
\caption{\label{figure_witnesses}
A sketch of the set of block positive operators (entanglement witnesses) 
for $\Hspdimension=2$. It includes the set of positive operators (quantum states) 
and the set of separable states. Here, any entanglement witness $W$ belonging 
to the border of $\bposops$ is optimal ($\omega$ is the line dual to $W$). 
Three of the optimal witnesses ($W_1$, $W_2$, $W_3$) 
are extreme points of $\bposops$ and the corresponding 
dual lines ($\omega_1$, $\omega_2$, $\omega_3$) determine completely
the shape of the set of separable states in this plot.}
\end{center}
\end{figure}

By modifying Figure \ref{figuredualities1} a little, we get 
a sketch that illustrates the important notion of an 
\textit{optimal entanglement witness} \cite{LKCH00} (cf. also \cite{KABLA08}). 
By definition, a block positive operator 
$W\in\linearopsb{\HilbertSp\otimes\HilbertSp}$ 
is called \textit{optimal} if and only if the set 
$\Delta_W:=\left\{\rho\in\posopsb{\HilbertSp}|\Trb{\rho W}<0\right\}$ 
is maximal (with respect of inclusion) within the family
 of sets $\Delta_{W'}$ (for $W'\in\bposopsb{\HilbertSp\otimes\HilbertSp}$). 
It is known \cite{LKCH00} that optimal witnesses have to lie 
on the boundary of $\bposopsb{\HilbertSp\otimes\HilbertSp}$ 
and in the case of Figure \ref{figure_witnesses} (page \pageref{figure_witnesses}) 
every element of the boundary of the triangle representing $\bposopsb{\HilbertSp\otimes\HilbertSp}$ 
corresponds to an optimal entanglement witness. 
The picture suggests that not every optimal witness is needed to determine the shape of the
 set of separable states, $\sepopsb{\HilbertSp\otimes\HilbertSp}=\dualityops{\bposopsb{\HilbertSp\otimes\HilbertSp}}$. 
Indeed, it is possible to consider only the optimal witnesses which are extreme points of the intersection
 of $\bposopsb{\HilbertSp\otimes\HilbertSp}$ with the hyperplane $\Tra{W}=1$. This is so because 
we have the following propositions,\footnote{The two propositions are not used in later parts of the paper, but they make for an apt comment to Figure \ref{figure_witnesses}. In spite of their simple character, they seem not to be fully realized by all scientists working in the field.}
\begin{proposition}\label{extremalpointsBP}
 An operator $\rho\in\linearopsb{\HilbertSp\otimes\HilbertSp}$ is separable iff $\Trb{W\rho}\geqslant 0$ for 
all $W$ extreme in 
$\bposopsb{\HilbertSp\otimes\HilbertSp}':=\left\{W\in\bposopsb{\HilbertSp\otimes\HilbertSp}|\Tra{W}=1\right\}$.
\begin{proof}
 The ``only if'' part is obvious from Proposition \ref{ksposdualkpos}. 
Let $\textnormal{e}\bposopsb{\HilbertSp\otimes\HilbertSp}'$ denote the set of 
extreme points of $\bposopsb{\HilbertSp\otimes\HilbertSp}'$. The ``if'' part of the 
proposition follows because $\bposopsb{\HilbertSp\otimes\HilbertSp}'=\convhull{\textnormal{e}\bposopsb{\HilbertSp\otimes\HilbertSp}'}$ 
as well as $\bposopsb{\HilbertSp\otimes\HilbertSp}=\setR^+_0\bposopsb{\HilbertSp\otimes\HilbertSp}'$, where the
 first equality is a consequence of the Krein-Milman theorem ($\bposopsb{\HilbertSp\otimes\HilbertSp}'$ is compact) 
and the latter holds because a block positive operator $W$ has zero trace only if $W=0$.
All in all, we get $\bposopsb{\HilbertSp\otimes\HilbertSp}=\setR^+_0\convhull{\textnormal{e}\bposopsb{\HilbertSp\otimes\HilbertSp}'}$ 
and the proposition follows from Proposition \ref{ksposdualkpos} by using the linearity of the trace.
\end{proof}
\end{proposition}

\begin{proposition}\label{everyextrisopt}
 Every extreme point of $\bposopsb{\HilbertSp\otimes\HilbertSp}'$ is an optimal entanglement witness.
\begin{proof}
 According to Theorem 1 in \cite{LKCH00}, an entanglement witness $W$ is 
optimal iff $\left(1+\varepsilon\right)W-\varepsilon P\not\in\bposopsb{\HilbertSp\otimes\HilbertSp}$
 for arbitrary $\varepsilon>0$ and a nonzero 
$P\in\posopsb{\HilbertSp\otimes\HilbertSp}$. Assume that $W$ is an extreme point in
 $\bposopsb{\HilbertSp\otimes\HilbertSp}'$ and 
$\left(1+\varepsilon\right)W-\varepsilon P\in\bposopsb{\HilbertSp\otimes\HilbertSp}$ for 
some $\varepsilon>0$, $P\in\posopsb{\HilbertSp\otimes\HilbertSp}\setminus\left\{0\right\}$. 
This is the same as $W-\xi P\in\bposopsb{\HilbertSp\otimes\HilbertSp}$ for some $\xi>0$ or
 $W-\upsilon P/\Tra{P}\in\bposopsb{\HilbertSp\otimes\HilbertSp}$ for some $\upsilon>0$. 
Then, of course, $W':=\left(1+\upsilon\right)W-\upsilon P/\Tra{P}$ is an element 
of $\bposopsb{\HilbertSp\otimes\HilbertSp}'$. But this contradicts extremality of $W$ 
since $W=W'/\left(1+\upsilon\right)+\upsilon P/\left(\left(1+\upsilon\right)\Tra{P}\right)$, 
$1/\left(1+\upsilon\right)+\upsilon/\left(1+\upsilon\right)=1$ and 
both $W'$ and $P/\Tra{P}$ are elements of $\bposopsb{\HilbertSp\otimes\HilbertSp}'$. 
Thus $\left(1+\varepsilon\right)W-\varepsilon P\not\in\bposopsb{\HilbertSp\otimes\HilbertSp}$
 for arbitrary $\varepsilon>0$ and 
$P\in\posopsb{\HilbertSp\otimes\HilbertSp}\setminus\left\{0\right\}$, so $W$ is optimal.
\end{proof}
\end{proposition}

It is therefore natural to define \textit{extreme entanglement witnesses} as the extreme points of $\bposopsb{\HilbertSp\otimes\HilbertSp}'$ and to give priority to witnesses which are not only optimal, but also extreme. We have
\begin{equation}\label{extremaldef}
{\mathbf{\textnormal{\bf extreme entanglement witnesses}
=\textnormal{\bf extreme points of }\bposopsb{\HilbertSp\otimes\HilbertSp}'}},\nonumber
\end{equation}
and in principle, no other witnesses are needed to describe the
set of separable states.

It should be kept in mind that Fig.  \ref{figure_witnesses} presents a highly simplified sketch of the problem.
Even in the simplest
possible case of a $2 \times 2$ system
the set of separable states is $15$ dimensional
and it is well known that this convex set is not a polytope
and its geometry is rather involved \cite{BZ06}. It is not our intention to discuss it here in detail and we return to the subject of duality relations. 

\vskip 3 mm\noindent Using the results presented earlier, it is straightforward to show the following
\begin{corollary}\label{kmdualities}Let $k,m$ be positive integers. We have
 $\dualitymaps{\kmdecmapsb{k}{m}{\HilbertSp}}=
\kmspositivitymapsb{k}{m}{\HilbertSp}$ and 
$\dualitymaps{\kmspositivitymapsb{k}{m}{\HilbertSp}}=
\kmdecmapsb{k}{m}{\HilbertSp}$
\qed
\end{corollary}

The next result, related  to composition properties of 
maps \cite{BZ06,CK07,MastersThesis},
 will be crucial for our later discussion
\begin{theorem}
\label{prodthmkSPkP}
$\kspositivitymapsb{k}{\HilbertSp}\circ\kpositivitymapsb{k}{\HilbertSp}=
\kpositivitymapsb{k}{\HilbertSp}\circ\kspositivitymapsb{k}{\HilbertSp}=
\kspositivitymapsb{k}{\HilbertSp}$
\begin{proof}
 Being more explicit, we want to prove that
 $\Phi\circ\Psi\in\kspositivitymapsb{k}{\HilbertSp}$ and 
$\Psi\circ\Phi\in\kspositivitymapsb{k}{\HilbertSp}$ for 
arbitrary $k\in\setN$, whenever $\Phi\in\kspositivitymapsb{k}{\HilbertSp}$ and 
$\Psi\in\kpositivitymapsb{k}{\HilbertSp}$. It is sufficient to show this for 
$\Phi=\Ad_a$ with an arbitrary $a\in\linearopsb{\HilbertSp}$ of rank $\leqslant k$. 
We prove first that $\Psi\circ\Ad_a$ is an element of 
$\kspositivitymapsb{k}{\HilbertSp}$. For this we shall need the following lemma
\begin{lemma}
\label{lemmaPhikposofdiad}
 Let $\Psi\in\linearmapsb{\HilbertSp}$ be $k$-positive. 
For any $k$-element set of vectors $\seq{\psi_i}{i=1}{k}$, 
there exists $m\in\setN$ and vectors $\seq{\xi^{\left(n\right)}_l}{l,n=1}{l=k,n=m}$ 
such that
\begin{equation}
\label{Phiofadiad}
 \Psi\left(\diad{\psi_i}{\psi_j}\right)=\sum_{n=1}^m\diad{\xi^{\left(n\right)}_i}{\xi^{\left(n\right)}_j}
\end{equation}
for all $i,j\in\left\{1,\ldots,k\right\}$.
\begin{proof}
The operator $\left[\Psi\left(\diad{\psi_i}{\psi_j}\right)\right]_{i,j=1}^k$ belongs to
 $\linearopsb{\setC^k\otimes\HilbertSp}$. Since $\psi$ is positive, 
$\left[\Psi\left(\diad{\psi_i}{\psi_j}\right)\right]\in\posopsb{\setC\otimes\HilbertSp}$, 
hence is a sum of positive rank $1$ operators, which are necessarily of the form 
$\left[\diad{\xi_i^{\left(n\right)}}{\xi_j^{\left(n\right)}}\right]_{i,j=1}^k$ with
 $\left\{\xi_l^{\left(n\right)}\right\}_{l,n=1}^{l=k,n=m}$ as in the statement of the theorem.
\end{proof}
\end{lemma}
Now we can prove that $\Psi\circ\Ad_a\in\kspositivitymapsb{k}{\HilbertSp}$. 
Let us take an arbitrary element $x\in\linearopsb{\HilbertSp}$. 
The fact that $\rk a\leqslant k$ is equivalent to 
$a=\sum_{i=1}^k\diad{\phi_i}{\psi_i}$ for some vectors $\seq{\phi_i}{i=1}{k},
\seq{\psi_j}{j=1}{k}\subset\HilbertSp$. Thus we get 
\begin{equation}
\label{formofAdmap}
 \Ad_a\left(x\right)=\sum_{i,j=1}^k\innerpr{\phi_i}{x\phi_j}
\diad{\psi_i}{\psi_j}.
\end{equation}
Now we calculate the action of $\Psi\circ\Ad_a$ on $x$,
\begin{equation}\label{actionofPhiAd}
 \left(\Psi\circ\Ad_a\right)x=\sum_{i,j=1}^k\innerpr{\phi_i}{x\phi_j}
\Psi\left(\diad{\psi_i}{\psi_j}\right)=\sum_{l=1}^m\sum_{i,j=1}^k\innerpr{\phi_i}
{x\phi_j}\diad{\xi^{\left(l\right)}_i}{\xi^{\left(l\right)}_j}.
\end{equation}
This is a sum of terms of the form \eqref{formofAdmap} and 
we get $\Psi\circ\Ad_a=\sum_{l=1}^m\Ad_{a_l}$, where the 
operators $a_l:=\sum_{j=1}^k\diad{\phi_j}{\xi^{\left(l\right)}_j}$ all 
have rank lower or equal $k$. Thus we have proved 
$\Psi\circ\Ad_a\in\kspositivitymapsb{k}{\HilbertSp}$, 
which implies that $\Psi\circ\Phi\in\kspositivitymapsb{k}{\HilbertSp}$ 
for arbitrary $\Phi\in\kspositivitymapsb{k}{\HilbertSp}$. 
We still need to show that $\Phi\circ\Psi\in\kspositivitymapsb{k}{\HilbertSp}$. 
This can be easily deduced from the following lemma,
\begin{lemma}\label{lemmatransposed}
 Let $\Phi$ be an element of $\kspositivitymapsb{k}{\HilbertSp}$ and $\Psi$ 
an element of $\kpositivitymapsb{k}{\HilbertSp}$.  Let $\hconj{\Phi}$, 
$\hconj{\Psi}$ be the adjoint operators of $\Phi$, $\Psi$ (resp.) 
with respect to the Hilbert-Schmidt product on $\linearopsb{\HilbertSp}$,
 given by the formula \eqref{HSprod} with $a,b\in\linearopsb{\HilbertSp}$.
 We have $\hconj{\Phi}\in\kspositivitymapsb{k}{\HilbertSp}$ and 
$\hconj{\Psi}\in\kpositivitymapsb{k}{\HilbertSp}$.
\begin{proof}
Just as $\posopsb{\HilbertSp\otimes\HilbertSp}$, the set 
$\posopsb{\setC^k\otimes\HilbertSp}$ is self-dual. Thus we have that
$x\in\posopsb{\setC^k\otimes\HilbertSp}\Leftrightarrow\Trb{\hconj{x}y}
\geqslant 0\,\forall_{y\in\posopsb{\setC^k\otimes\HilbertSp}}$. 
The definition of $k$-positivity of $\Psi$ can be restated as
\begin{equation}
\label{defkpos2}
 \Trb{\hconjb{\left(\identitymapn{k}\otimes\Psi\right) x}y}\geqslant 0\,
\forall_{x,y\in\posopsb{\setC^k\otimes\HilbertSp}}.
\end{equation}
Equivalently,
\begin{equation}\label{defkpos3}
 \Trb{\hconjb{\left(\identitymapn{k}\otimes\hconj{\Psi}\right) y}x}\geqslant0\,
\forall_{x,y\in\posopsb{\setC^k\otimes\HilbertSp}}.
\end{equation}
But this is just the condition \eqref{defkpos2} for $\hconj{\Psi}$. 
Hence $\Psi\in\kpositivitymapsb{k}{\HilbertSp}\Leftrightarrow\hconj{\Psi}\in
\kpositivitymapsb{k}{\HilbertSp}$. To prove an analogous equivalence for $\Phi$, 
it is enough to consider the specific case $\Phi=\Ad_a$ with $\rk a\leqslant k$. 
We have
\begin{equation}
\label{adjointPhi}
 \Trb{\hconjb{\Ad_a\left(x\right)}y}=\Trb{\hconjb{\hconj{a}xa}y}=
\Trb{\hconj{x}\hconjb{ay\hconj{a}}}=\Trb{x\hconjb{\Ad_{\hconj{a}}\left(y\right)}}
\end{equation}
This gives us $\hconjb{\Ad_a}=\Ad_{\hconj{a}}$. 
The ranks of $a$ and $\hconj{a}$ are equal, so 
$\Ad_a\in\kspositivitymapsb{k}{\HilbertSp}\Leftrightarrow\hconjb{\Ad_{a}}\in
\kspositivitymapsb{k}{\HilbertSp}$, which implies 
$\Phi\in\kspositivitymapsb{k}{\HilbertSp}\Leftrightarrow\hconj{\Phi}\in
\kspositivitymapsb{k}{\HilbertSp}$ and finishes the proof of the lemma.
\end{proof}
\end{lemma}
Now we can finish the proof of Theorem \ref{prodthmkSPkP}. 
By Lemma \ref{lemmatransposed}, $\Phi\circ\Psi\in\kspositivitymapsb{k}{\HilbertSp}$ 
is equivalent to $\hconjb{\Phi\circ\Psi}=\hconj{\Psi}\circ\hconj{\Phi}
\in\kspositivitymapsb{k}{\HilbertSp}$. The last equality holds according to
 Lemma \ref{lemmatransposed} and to the first part of the theorem.
\end{proof}
\end{theorem}
\noindent In short, we proved that for any $\Phi$ $k$-superpositive and 
$\Psi$ $k$-positive, the products $\Phi\circ\Psi$ and $\Psi\circ\Phi$ are 
$k$-superpositive.

It is good to notice that Theorem \ref{prodthmkSPkP} justifies 
the name \textit{entanglement breaking channels}, which is often used 
for superpositive, trace preserving maps of $\linearopsb{\HilbertSp}$. 
To make this precise, we show the following
\begin{corollary}Let $\Phi$ be superpositive. 
For any $\rho\in\posopsb{\HilbertSp\otimes\HilbertSp}$, we have
\begin{equation}
\label{eqPhirho}
\left(\identitymap\otimes\Phi\right)\rho\in\sepopsb{\HilbertSp\otimes\HilbertSp} 
\end{equation}
\begin{proof}
Since $\Jamiso{\cpositivitymapsb{\HilbertSp}}=\posopsb{\HilbertSp\otimes\HilbertSp}$,
 where $\Jamiosymb$ is the isomorphism defined in \eqref{Jamiolkowskimap},
 we have
\begin{equation}
 \label{rhoisequal}
\rho=\left(\identitymap\otimes\Psi\right)\proj{\psi_+}
\end{equation}
for a suitably chosen $\Psi\in\cpositivitymapsb{\HilbertSp}$. We have
\begin{equation}
 \label{idotimesphiidotimespsi}
\left(\identitymap\otimes\Phi\right)\rho=\left(\identitymap\otimes\Phi\right)
\left(\identitymap\otimes\Psi\right)\proj{\psi_+}=\left(\identitymap\otimes\Phi
\circ\Psi\right)\proj{\psi_+}.
\end{equation}
Because $\cpositivitymapsb{\HilbertSp}$ is a subset of $\positivitymapsb{\HilbertSp}$, 
$\Psi$ is an element of $\positivitymapsb{\HilbertSp}$ an we get from Theorem \ref{prodthmkSPkP} 
the inclusion $\Phi\circ\Psi\in\spositivitymapsb{\HilbertSp}$. By 
Proposition \ref{propkSPkSep}, the operator 
$\left(\identitymap\otimes\Phi\circ\Psi\right)\proj{\psi_+}$ is separable. 
Comparing this with \eqref{idotimesphiidotimespsi}, we immediately see that 
\eqref{eqPhirho} is true.\end{proof}
\end{corollary}
Obviously, it is possible to repeat the argument given above in the case when we 
assume $k$-superpositivity of $\Phi$ and demand $k$-separability of 
$\left(\identitymap\otimes\Phi\right)\rho$.
Therefore one could think of calling $k$-superpositive and trace 
preserving maps \textit{$k$-separability inducing channels}.

We shall finish this section with a number of characterizations of the sets
 $\kspositivitymapsb{k}{\HilbertSp}$ and $\kpositivitymapsb{k}{\HilbertSp}$. 
Together with Theorem \ref{prodthmkSPkP}, the following four theorems should 
be regarded as some of the most important material included in the paper and 
be studied with care.
\begin{theorem}\label{characterizationsksp}
Let $\Phi\in\hermicitymapsb{\HilbertSp}$ and $k\in\setN$. The following 
conditions are equivalent:
\begin{enumerate}[1)]
\item $\Phi\in\kspositivitymapsb{k}{\HilbertSp}$,
\item $\Psi\circ\Phi\in\kspositivitymapsb{k}{\HilbertSp}\,
\forall_{\Psi\in\kpositivitymapsb{k}{\HilbertSp}}$,
\item $\Psi\circ\Phi\in\cpositivitymapsb{\HilbertSp}\,
\forall_{\Psi\in\kpositivitymapsb{k}{\HilbertSp}}$,
\item $\Trb{\proj{\psi_+}\left(\identitymap\otimes
\left(\Psi\circ\Phi\right)\right)\left(\proj{\psi_+}\right)}\geqslant 0\,\,
\forall_{\Psi\in\kpositivitymapsb{k}{\HilbertSp}}$.
\end{enumerate}
\begin{proof}
$1)\Rightarrow 2)$ As we know from Theorem \ref{prodthmkSPkP}, 
$\Psi\circ\Phi\in\kspositivitymapsb{k}{\HilbertSp}$ for 
$\Psi\in\kpositivitymapsb{k}{\HilbertSp}$ and $\Phi\in\kspositivitymapsb{k}
{\HilbertSp}$. This proves 2)

\noindent $2)\Rightarrow 3)$ This implication is obvious because 
$\kspositivitymapsb{k}{\HilbertSp}\subset\kpositivitymapsb{k}{\HilbertSp}$

\noindent $3)\Rightarrow 4)$ We know from $3)$ that $\Psi\circ\Phi$ is 
completely positive. As a consequence of Choi's theorem (Proposition \ref{Choithm}),
 $\Choimatr{\Psi\circ\Phi}=\left(\identitymap\otimes\left(\Psi\circ\Phi\right)\right)
\left(\proj{\psi_+}\right)$ is positive. Thus we have $\Trb{\proj{\psi_+}
\Choimatr{\Psi\circ\Phi}}\geqslant 0$, which is precisely the statement in $4)$. 

\noindent $4)\Rightarrow 1)$ Let $\Theta_{\Psi,\Phi}$ denote 
$\Trb{\proj{\psi_+}\left(\identitymap\otimes\left(\Psi\circ\Phi\right)\right)
\left(\proj{\psi_+}\right)}$. We calculate
\begin{eqnarray}
\label{Erlinglemma311}
\Theta_{\Psi,\Phi}
=&\Trb{\proj{\psi_+}\left(\identitymap\otimes\Psi\right)\circ\left(\identitymap\otimes\Phi\right)\left(\proj{\psi_+}\right)}&=\\
=&\Trb{\hconjb{\identitymap\otimes\Psi}\left(\proj{\psi_+}\right)\left(\identitymap\otimes\Phi\right)\left(\proj{\psi_+}\right)}&=\nonumber\\
=&\Trb{\left(\identitymap\otimes\hconj{\Psi}\right)\left(\proj{\psi_+}\right)\left(\identitymap\otimes\Phi\right)\left(\proj{\psi_+}\right)}&=\Trb{\Choimatr{\hconj{\Psi}}\Choimatr{\Phi}}.\nonumber
\end{eqnarray}
Thus the condition
 $\Theta_{\Psi,\Phi}\geqslant 0\, \forall_{\Psi\in\kpositivitymaps{\HilbertSp}}$, 
which we have in $4)$, is the same as
\begin{equation}
  \label{pointfourduality1}
 \Trb{\Choimatr{\hconj{\Psi}}\Choimatr{\Phi}}\geqslant 0\,
\forall_{\Psi\in\kpositivitymapsb{k}{\HilbertSp}}
\end{equation}
Using Lemma \ref{lemmatransposed} again, 
we see that \eqref{pointfourduality1} is equivalent to
\begin{equation}\label{pointfourduality2}
\Trb{\Choimatr{\Psi}\Choimatr{\Phi}}\geqslant 0\,
\forall_{\Psi\in\kpositivitymapsb{k}{\HilbertSp}}.
\end{equation}
Comparing this with the definition \eqref{defdualconeinLH} 
of the dual cone of $\kpositivitymapsb{k}{\HilbertSp}$ and 
using Proposition \ref{ksposdualkpos}, we obtain
\begin{equation}\label{PhiinkSP}
 \Phi\in\dualitymapsb{\kpositivitymapsb{k}{\HilbertSp}}=
\kspositivitymapsb{k}{\HilbertSp},
\end{equation}
which is $1)$.
\end{proof}
\end{theorem}
The following three characterization theorems can be 
proved in practically the same way as Theorem \ref{characterizationsksp}.

\begin{theorem}\label{characterizationsksp2}
Let $\Phi\in\hermicitymapsb{\HilbertSp}$ and $k\in\setN$. 
The following conditions are equivalent:
\begin{enumerate}[1)]
\item $\Phi\in\kspositivitymapsb{k}{\HilbertSp}$,
\item $\Phi\circ\Psi\in\kspositivitymapsb{k}{\HilbertSp}\,
\forall_{\Psi\in\kpositivitymapsb{k}{\HilbertSp}}$,
\item $\Phi\circ\Psi\in\cpositivitymapsb{\HilbertSp}\,
\forall_{\Psi\in\kpositivitymapsb{k}{\HilbertSp}}$,
\item $\Trb{\proj{\psi_+}\left(\identitymap\otimes\left(\Phi\circ\Psi\right)\right)
\left(\proj{\psi_+}\right)}\geqslant 0\,\,
\forall_{\Psi\in\kpositivitymapsb{k}{\HilbertSp}}$.
\end{enumerate}
\qed
\end{theorem}

\begin{theorem}\label{characterizationskp1}
Let $\Phi\in\hermicitymapsb{\HilbertSp}$ and $k\in\setN$. The following conditions are equivalent:
\begin{enumerate}[1)]
\item $\Phi\in\kpositivitymapsb{k}{\HilbertSp}$,
\item $\Psi\circ\Phi\in\kspositivitymapsb{k}{\HilbertSp}\,
\forall_{\Psi\in\kspositivitymapsb{k}{\HilbertSp}}$,
\item $\Psi\circ\Phi\in\cpositivitymapsb{\HilbertSp}\,
\forall_{\Psi\in\kspositivitymapsb{k}{\HilbertSp}}$,
\item $\Trb{\proj{\psi_+}\left(\identitymap\otimes
\left(\Psi\circ\Phi\right)\right)\left(\proj{\psi_+}\right)}\geqslant 0\,\,
\forall_{\Psi\in\kspositivitymapsb{k}{\HilbertSp}}$.
\end{enumerate}
\qed
\end{theorem}

\begin{theorem}\label{characterizationskp2}
Let $\Phi\in\hermicitymapsb{\HilbertSp}$ and $k\in\setN$. The following conditions 
are equivalent:
\begin{enumerate}[1)]
\item $\Phi\in\kpositivitymapsb{k}{\HilbertSp}$,
\item $\Phi\circ\Psi\in\kspositivitymapsb{k}{\HilbertSp}\,
\forall_{\Psi\in\kspositivitymapsb{k}{\HilbertSp}}$,
\item $\Phi\circ\Psi\in\cpositivitymapsb{\HilbertSp}\,
\forall_{\Psi\in\kspositivitymapsb{k}{\HilbertSp}}$,
\item $\Trb{\proj{\psi_+}\left(\identitymap\otimes\left(\Phi\circ\Psi\right)\right)
\left(\proj{\psi_+}\right)}\geqslant 0\,\,
\forall_{\Psi\in\kspositivitymapsb{k}{\HilbertSp}}$.
\end{enumerate}
\qed
\end{theorem}
Theorem \ref{characterizationsksp2} is much the same as Theorem 
\ref{characterizationsksp}, but the order of the operators $\Psi$, 
$\Phi$ is different in these theorems. Theorems \ref{characterizationskp1} 
and \ref{characterizationskp2} are in complete analogy with 
\ref{characterizationsksp} and \ref{characterizationsksp2} (resp.), 
but the roles of $k$-positive and $k$-superpositive maps have been exchanged. 
In section 4 we shall add two more to the list of equivalent conditions in the the above theorems, see Corollaries \ref{cor42Erling} and \ref{cor43Erling}.

We should remark that the four theorems given above make up a 
broad generalization of a number of relatively well known facts 
about the sets $\positivitymapsb{\HilbertSp}$, $\cpositivitymapsb{\HilbertSp}$ and 
$\spositivitymapsb{\HilbertSp}$,
\begin{eqnarray}
 \label{conditionsSP}
\Phi\in\spositivitymapsb{\HilbertSp}&\Longleftrightarrow&
\Psi\circ\Phi\in\cpositivitymapsb{\HilbertSp}\,
\forall_{\Psi\in\positivitymapsb{\HilbertSp}}\\
\label{conditionsCP}
\Phi\in\cpositivitymapsb{\HilbertSp}&\Longleftrightarrow&
\Psi\circ\Phi\in\cpositivitymapsb{\HilbertSp}\,
\forall_{\Psi\in\cpositivitymapsb{\HilbertSp}}\\
\label{conditionsP}
\Phi\in\positivitymapsb{\HilbertSp}&\Longleftrightarrow&
\Psi\circ\Phi\in\cpositivitymapsb{\HilbertSp}\,
\forall_{\Psi\in\spositivitymapsb{\HilbertSp}}
\end{eqnarray}
(these can be found on page 345 of \cite{BZ06}). We should emphasize that the results like \eqref{conditionsSP}-\eqref{conditionsP} 
and our four theorems do not simply follow from the closedness relations of the type 
$\Phi,\Psi\in\cpositivitymapsb{\HilbertSp}\Rightarrow\Phi\circ\Psi\in\cpositivitymapsb{\HilbertSp}$ 
(and similarly for $\positivitymapsb{\HilbertSp}$, $\kpositivitymapsb{k}{\HilbertSp}$, $\kspositivitymapsb{k}{\HilbertSp}$ 
and $\spositivitymapsb{\HilbertSp}$).


\section{Mapping cones}\label{secmappingcones}

In the previous sections we have studied maps of $\linearopsb{\HilbertSp}$ into
itself for $\HilbertSp$ a finite dimensional Hilbert space, and much of the technical
 work has involved the Choi matrix \eqref{Choimatrix} and the Jamio\l kowski
 \eqref{Jamiolkowskimap} isomorphism. In more general situations these techniques are
 not available, and one of us introduced in \cite{St86} an alternative approach to
 study positivity properties of maps of a \cstardollar-algebra into
 $\linearopsb{\HilbertSp}$. We now recall some of the definitions. For simplicity we
 continue to assume $\HilbertSp$ is finite dimensional.

Let $A$ be a \cstardollar-algebra. Then there is a duality between bounded linear
 maps $\Phi$ of $A$ into $\linearopsb{\HilbertSp}$ and linear functionals
 $\tilde\Phi$ on $A\otimes\linearopsb{\HilbertSp}$ given by
\begin{equation}\label{phitilde}
 \tilde\Phi\left(a\otimes b\right)=\Trb{\Phi\left(a\right)b^{\transpos}},\,a\in
 A,\,b\in\linearopsb{\HilbertSp},
\end{equation}
where $\Tra{}$ is the usual trace on $\linearopsb{\HilbertSp}$ and $\transpos$ the
 transpose. Furthermore, $\Phi$ is positive iff $\tilde\Phi$ is positive on the cone
 $A^+\otimes\posopsb{\HilbertSp}$ of separable operators. We say a nonzero cone
 $\mathcal K$ in $\positivitymapsb{\HilbertSp}$ is a \textit{mapping cone} if
 $\Phi\in\mathcal{K}$ implies $\Psi\circ\Phi\circ\Upsilon\in\mathcal{K}$ for all
 $\Psi,\Upsilon\in\cpositivitymapsb{\HilbertSp}$. Well known examples are
 $\positivitymapsb{\HilbertSp}$, $\cpositivitymapsb{\HilbertSp}$, the copositive maps
 and $\spositivitymapsb{\HilbertSp}$. We define
\begin{equation}\label{defPAK}
 \PAK{A}{\mathcal{K}}:=\left\{x\in A\otimes\linearopsb{\HilbertSp}|x=\hconj{x},\identitymap\otimes\Psi\left(x\right)
\geqslant 0\,\forall_{\Psi\in\mathcal{K}}\right\},
\end{equation}
where $\identitymap$ denotes the identity map on $\linearmapsb{\HilbertSp}$,
 $\PAK{A}{\mathcal{K}}$ is a proper closed cone in $A\otimes\linearopsb{\HilbertSp}$
 containing the cone $A^+\otimes\posopsb{\HilbertSp}$.

We say $\Phi$ is \textit{$\mathcal K$-positive} if $\tilde\Phi$ is positive on
$\PAK{A}{\mathcal{K}}$, and denote by $\Kapositivemapsb{\mathcal K}{\HilbertSp}$ the
set of $\mathcal K$-positive maps of $A$ into $\linearopsb{\HilbertSp}$. Then $\Phi$
 is completely positive iff $\Phi$ is $\cpositivitymapsb{\HilbertSp}$-positive
 \cite[Theorem 3.2]{St86} iff $\tilde\Phi$ is a positive linear functional on
 $A\otimes\linearopsb{\HilbertSp}$.

If $A$ is contained in a larger \cstardollar-algebra $B$ then $\mathcal K$-positive maps from $A$ to $\linearopsb{H}$ have $\mathcal K$-positive extensions to maps from $B$ into $\linearopsb{\HilbertSp}$, \cite[Theorem 3.1]{St86}. In particular, this holds if $B=\linearopsb{\HilbertSp}$. Therefore the results from the previous sections are applicable in much more general situations as soon as we can show $\mathcal{K}=\Kapositivemapsb{\mathcal K}{\HilbertSp}$. The main results in the present section are concerned with this problem, and we shall show that it has an affirmative solution for the cones $\kpositivitymapsb{k}{\HilbertSp}$, $\kspositivitymapsb{k}{\HilbertSp}$ and leave the discussion of $\kmdecmapsb{k}{m}{\HilbertSp}$ and $\kmspositivitymapsb{k}{m}{\HilbertSp}$ to the reader.
\begin{lemma}\label{lemma41Erling}
 The cones $\kpositivitymapsb{k}{\HilbertSp}$, $\kspositivitymapsb{k}{\HilbertSp}$, $\kmdecmapsb{k}{m}{\HilbertSp}$ and $\kmspositivitymapsb{k}{m}{\HilbertSp}$ are all mapping cones.
\begin{proof}
 If $\Phi\in\kpositivitymapsb{k}{\HilbertSp}$ then $\identitymapn{k}\otimes\Phi\geqslant 0$, where $\identitymapn{k}$ is the identity map on a $k$-dimensional Hilbert space. Thus if $\Psi\in\cpositivitymapsb{\HilbertSp}$,
\begin{equation}\label{phialpha1}
\identitymapn{k}\otimes\left(\Phi\circ\Psi\right)=\left(\identitymapn{k}\otimes\Phi\right)\left(\identitymapn{k}\otimes\Psi\right)\geqslant 0
\end{equation}
and
\begin{equation}\label{phialpha2}
 \identitymapn{k}\otimes\left(\Psi\circ\Phi\right)=\left(\identitymapn{k}\otimes\Psi\right)\left(\identitymapn{k}\otimes\Phi\right)\geqslant 0.
\end{equation}
Thus $\kpositivitymapsb{k}{\HilbertSp}$ is a mapping cone.
\end{proof}
\end{lemma}

If $\rk a\leqslant k$ then for all $b\in\linearopsb{\HilbertSp}$, $\rk ab\leqslant k$ and $\rk ba\leqslant k$. Thus $\Ad_{b}\circ\Ad_{a}=\Ad_{ba}\in\kspositivitymapsb{k}{\HilbertSp}$, and $\Ad_{a}\circ\Ad_{b}\in\kspositivitymapsb{k}{\HilbertSp}$. It follows that $\kspositivitymapsb{k}{\HilbertSp}$ is a mapping cone. From the definitions of $\kmdecmapsb{k}{m}{\HilbertSp}$ and $\kmspositivitymapsb{k}{m}{\HilbertSp}$ it follows that they are also mapping cones.
\begin{theorem}\label{thm42Erling}
 $\kspositivitymapsb{k}{\HilbertSp}=\Kapositivemapsb{\kspositivitymapsb{k}{\HilbertSp}}{\HilbertSp}$, and $\kpositivitymapsb{k}{\HilbertSp}=\Kapositivemapsb{\kpositivitymapsb{k}{\HilbertSp}}{\HilbertSp}$.
\begin{proof}
By Theorem \ref{characterizationskp1}, $\Phi\in\kpositivitymapsb{k}{\HilbertSp}$ iff $\Psi\circ\Phi\in\cpositivitymapsb{\HilbertSp}$ for all $\Psi\in\kspositivitymapsb{k}{\HilbertSp}$. Hence by \cite[Theorem 1]{St09}, $\Phi\in\kpositivitymapsb{k}{\HilbertSp}$ iff $\Phi$ belongs to the dual cone $\dualitymaps{\Kapositivemapsb{\kspositivitymapsb{k}{\HilbertSp}}{\HilbertSp}}$ of $\Kapositivemapsb{\kspositivitymapsb{k}{\HilbertSp}}{\HilbertSp}$. By Proposition \ref{kposdualksp},
 $\kpositivitymapsb{k}{\HilbertSp}=\dualitymaps{\kspositivitymapsb{k}{\HilbertSp}}$. Thus $\kspositivitymapsb{k}{\HilbertSp}=\dualitymaps{\kpositivitymapsb{k}{\HilbertSp}}=\Kapositivemapsb{\kspositivitymapsb{k}{\HilbertSp}}{\HilbertSp}^{\circ\circ}=\Kapositivemapsb{\kspositivitymapsb{k}{\HilbertSp}}{\HilbertSp}$, proving the first statement.

Similarly by Proposition \ref{ksposdualkpos}, $\Phi\in\dualitymaps{\kpositivitymapsb{k}{\HilbertSp}}$ iff $\Phi\in\kspositivitymapsb{k}{\HilbertSp}$. Thus by Theorem \ref{characterizationsksp}, $\Phi\in\dualitymaps{\kpositivitymapsb{k}{\HilbertSp}}$ iff $\Psi\circ\Phi\in\cpositivitymapsb{\HilbertSp}$ for all $\Psi\in\kpositivitymapsb{k}{\HilbertSp}$, hence by Theorem 1 in \cite{St09} iff $\Phi\in\dualitymaps{\Kapositivemapsb{\kpositivitymapsb{k}{\HilbertSp}}{\HilbertSp}}$. Thus $\kpositivitymapsb{k}{\HilbertSp}=\Kapositivemapsb{\kpositivitymapsb{k}{\HilbertSp}}{\HilbertSp}$.
\end{proof}
\end{theorem}

Using the above theorem and its proof together with Theorem 1 in \cite{St09} we can add two more conditions to the equivalent conditions in Theorems \ref{characterizationsksp} and \ref{characterizationskp1},
\begin{corollary}\label{cor42Erling}
 The following conditions are equivalent for $\Phi\in\hermicitymapsb{\HilbertSp}$,
\begin{enumerate}[1)]
\item $\Phi\in\kpositivitymapsb{k}{\HilbertSp}$, i.e. $\Phi$ is $k$-positive,
\item $\identitymap\otimes\Psi\left(\Choimatr{\Phi}\right)\geqslant 0\,\forall_{\Psi\in\kspositivitymapsb{k}{\HilbertSp}}$,
\item $\tilde\Phi\circ\left(\identitymap\otimes\Psi\right)\geqslant 0\,\forall_{\Psi\in\kspositivitymapsb{k}{\HilbertSp}}$.
\end{enumerate}
\qed
\end{corollary}

\begin{corollary}\label{cor43Erling}
 The following conditions are equivalent for $\Phi\in\hermicitymapsb{\HilbertSp}$,
\begin{enumerate}[1)]
\item $\Phi\in\kspositivitymapsb{k}{\HilbertSp}$, i.e. $\Phi$ is $k$-superpositive,
\item $\identitymap\otimes\Psi\left(\Choimatr{\Phi}\right)\geqslant 0\,\forall_{\Psi\in\kpositivitymapsb{k}{\HilbertSp}}$,
\item $\tilde\Phi\circ\left(\identitymap\otimes\Psi\right)\geqslant 0\,\forall_{\Psi\in\kpositivitymapsb{k}{\HilbertSp}}$.
\end{enumerate}
\qed
\end{corollary}
Using Proposition \ref{propkSPkSep}, it becomes evident that the condition 2) in Corollary \ref{cor43Erling} is the same as the $k$-positive maps criterion by Terhal and Horodecki \cite{TH00} (for $k=1$, we get the well known positive maps criterion by Horodeccy\footnote{Horodeccy is the plural form of the name Horodecki.}, \cite{HHH96a}). Corollary \ref{cor42Erling} provides us with an analogous characterization of the set of $k$-block positive operators: \textit{An operator $a\in\linearopsb{\HilbertSp\otimes\HilbertSp}$ is $k$-block positive iff $\left(\identitymap\otimes\Psi\right)a\geqslant 0$ for all $k$-superpositive maps $\Psi$}. 

Furthermore, the main theorem in \cite{Cl05} is a version of Corollary \ref{cor42Erling}, slightly modified to encompass $2$-copositive maps. One can easily deduce from it that the set of one-undistillable states on $\HilbertSp\otimes\HilbertSp$ is precisely $\kbposopsb{2}{\HilbertSp\otimes\HilbertSp}$.


\section{Concluding remarks}

In this paper we studied the structure of the set of positive maps from the space $\linearopsb{\HilbertSp}$ of linear operators on a finite-dimensional Hilbert space $\HilbertSp$ into itself. This topic is of substantial interest in quantum physics, since positive maps are closely related to the separability problem due to the positive maps criterion by Horodeccy \cite{HHH96a}. More generally, but less acute, positive maps are related to the separability problem because they correspond to hyperplanes that separate entangled states from the separable ones.

Here we developed general methods for proving results like the Horodeccy criterion, both in the situation where the Jamio{\l}kowski isomorphism is at hand and within a more general setup, where other techniques need to be used, based on mapping cones (cf. Section \ref{secmappingcones}). Our discussion concentrated on $k$-positive maps and on the dual cones of $k$-superpositive maps, consisting of completely positive maps that admit a Kraus representation by operators of rank $\leqslant k$ (such maps are also called \textit{partially entanglement breaking channels}, \cite{CK07}). We gave a number of characterization theorems (Theorems \ref{characterizationsksp}, \ref{characterizationskp1}, \ref{characterizationsksp2}, \ref{characterizationskp2} and Corollaries \ref{cor42Erling}, \ref{cor43Erling}) for both $k$-positive and $k$-superpositive maps, pertaining to their properties under taking compositions. Central to these results is the observation that a product of a $k$-superpositive map and a $k$-positive map is again a $k$-superpositive map (Theorem \ref{prodthmkSPkP}). We have not seen that particular result anywhere in the literature. Also our characterization theorems seem to appear for the first time in this paper.

We introduced (similary to \cite{CK07}, only using different notation) the cones of $\left(k,m\right)$-separable, $\left(k,m\right)$-decomposable and $\left(k,m\right)$-positive maps ($\kmspositivitymapsb{k}{m}{\HilbertSp}$, $\kmdecmapsb{k}{m}{\HilbertSp}$ and $\kmpositivitymapsb{k}{m}{\HilbertSp}$, respectively). The main results of this paper can be trivially generalized to these families of maps.

Most of our work relied on the simple and fine idea of \emph{duality} between convex cones \cite{Rockafellar}, which is nevertheless hard to grasp intuitively for spaces of dimension higher than $3$ (it is not even completely trivial for three-dimensional cones, see Figure \ref{figuredualities1}). We hope that the figures we included in Section \ref{sectionRelations} could help the reader to develop basic intuitions about the geometric background to our work. On that occasion we touched upon the question of optimality of entanglement witnesses. By pointing out that the extreme points of the set of unital witnesses are optimal, we tried to spill the idea that future efforts could concentrate on witnesses which are not only optimal, but also extreme.\footnote{We saw that optimal witnesses are closely related to points on the border of $\bposopsb{\HilbertSp\otimes\HilbertSp}$. It follows from the discussion in \cite{LKCH00} that they must lie there}

Within this paper several results by other authors \cite{Ja72,Cho75a,HHH96a,TH00,RA07,Cl05} appear as special cases of general theorems. Presented in the way we did it, they start to reveal a mathematical structure of a certain degree of generality. For a mathematican, it is natural to ask if there are many examples of this structure, or maybe it is very specific to the studied cones. In other words, the question is, how many are there interesting examples of mapping cones $\mathcal K$ in $\linearmapsb{\HilbertSp}$ such that $\Kapositivemapsb{\mathcal K}{\HilbertSp}=\mathcal{K}$? We do not know the answer at the moment. From a physicist's perspective, the key question here is to what extend the families $\kmspositivitymapsb{k}{m}{\HilbertSp}$, $\kmdecmapsb{k}{m}{\HilbertSp}$ and $\kmpositivitymapsb{k}{m}{\HilbertSp}$ can be useful in entanglement research and how our theorems can be applied in practice. The example of the paper \cite{Cl05} suggests that our discussion is not purely abstract and may relate to physically relevant questions
like the distillability of entanglement. 



\section{Acknowledgements}

It is a pleasure to thank A. Ac{\'i}n, D. Chru{\'s}ci{\'n}ski,  J. Korbicz, 
M. Lewenstein and P. Horodecki for helpful discussions.
We acknowledge financial support by the 
Polish Research Network LFPPI and the European grant COCOS.

\end{document}